\newif\ifShowKeys
\ShowKeystrue
\ShowKeysfalse

\documentclass[11pt,a4paper]{article} 
\pdfoutput=1
\usepackage[no-natbib-sort]{my-jheppub}

\usepackage{amsmath, amssymb}

\usepackage{bm}
\usepackage{environ}
\usepackage{mathrsfs}
\usepackage{array,arydshln}
\usepackage{booktabs}

\usepackage{graphicx,epsfig}
\usepackage{epic}

\usepackage{tensor} 							

\usepackage{float}
\usepackage[dvipsnames]{xcolor}
\definecolor{maroon}{rgb}{0.8,0.3,0.}

\usepackage{slashed}
\usepackage[nodayofweek]{date time}

\usepackage{hyperref}

\usepackage{aurical}
\usepackage[T1]{fontenc}

\usepackage{framed}
\definecolor{shadecolor}{RGB}{255, 230, 204}

\allowdisplaybreaks

\newmuskip\pFqmuskip

\newcommand*\pFq[6][8]{%
  \begingroup 
  \pFqmuskip=#1mu\relax
  \mathcode`\,=\string"8000
  \begingroup\lccode`\~=`\,
  \lowercase{\endgroup\let~}\pFqcomma
  {}_{#2}F_{#3}{\left[\genfrac..{0pt}{}{#4}{#5};#6\right]}%
  \endgroup
}

\newcommand*\pFtildeq[6][8]{%
  \begingroup 
  \pFqmuskip=#1mu\relax
  \mathcode`\,=\string"8000
  \begingroup\lccode`\~=`\,
  \lowercase{\endgroup\let~}\pFqcomma
  {}_{#2}\widetilde{F}_{#3}{\left[\genfrac..{0pt}{}{#4}{#5};#6\right]}%
  \endgroup
}

 \newcommand{\rf}[1]{(\ref{#1})}

\newcommand{\pFqcomma}{\mskip\pFqmuskip}

\makeatletter
\DeclareFontFamily{OMX}{MnSymbolE}{}
\DeclareSymbolFont{MnLargeSymbols}{OMX}{MnSymbolE}{m}{n}
\SetSymbolFont{MnLargeSymbols}{bold}{OMX}{MnSymbolE}{b}{n}
\DeclareFontShape{OMX}{MnSymbolE}{m}{n}{
    <-6>  MnSymbolE5
   <6-7>  MnSymbolE6
   <7-8>  MnSymbolE7
   <8-9>  MnSymbolE8
   <9-10> MnSymbolE9
  <10-12> MnSymbolE10
  <12->   MnSymbolE12
}{}
\DeclareFontShape{OMX}{MnSymbolE}{b}{n}{
    <-6>  MnSymbolE-Bold5
   <6-7>  MnSymbolE-Bold6
   <7-8>  MnSymbolE-Bold7
   <8-9>  MnSymbolE-Bold8
   <9-10> MnSymbolE-Bold9
  <10-12> MnSymbolE-Bold10
  <12->   MnSymbolE-Bold12
}{}

\let\llangle\@undefined
\let\rrangle\@undefined
\DeclareMathDelimiter{\llangle}{\mathopen}%
                     {MnLargeSymbols}{'164}{MnLargeSymbols}{'164}
\DeclareMathDelimiter{\rrangle}{\mathclose}%
                     {MnLargeSymbols}{'171}{MnLargeSymbols}{'171}
\makeatother

\newcommand{\be}{\begin{equation}}
\newcommand{\ee}{\end{equation}}

\newcommand{\mc}{\mathcal }

\newcommand{\bv}{\begin{verbatim}}
\newcommand{\ev}{\end{verbatim}}

\newcommand{\la}{\label}
\newcommand{\eps}{\varepsilon}

\newcommand{\vp}{\varphi}

\newcommand{\tr}{\text{tr}\,}

\def \ov {\over}
 \def \la {\label} \def \te {\textstyle} \def \b {\beta}
 \def \foot {\footnote}  

\def \ed {\end{document}}
\def \iffa {\iffalse}

\def \ov {\over}\def \te  {\textstyle} 
\def \ci {\cite}
\def \foot {\footnote}
\def \N {{\cal N}}
\def \b{\beta}
\def \m {\mu}
\def \n {\nu}
\def \del{\partial}

\def \k {\kappa}
\def \l {\lambda}
\def\z{\zeta}
\def \iffa {\iffalse}
 \def \no {\notag}
\def \a  {\alpha} 

\newcommand{\s}{\textrm{s}}
\newcommand{\Z}{\mc Z}
\newcommand{\tsusy}{\text{susy}}

\title{Superconformal index of higher derivative\\ $\mc N=1$  multiplets in 
four dimensions}

\author[a]{Matteo Beccaria} 
\author[b]{ and \ \ Arkady A. Tseytlin\footnote{Also at Lebedev Institute, Moscow.}} 

\abstract{Supersymmetric partition function 
 of $\mc N=1$ superconformal theories on $S^1_{\b} \times S^3$  is related 
to  the superconformal index receiving contributions from  short  representations.
The leading  coefficients  in the small $\b$ (high "temperature")    expansion of the index 
were previously related  to the  conformal  anomaly coefficients  of the  theory. 
Assumptions  underlying  universality  of these relations were tested only for simplest low-spin unitary 
multiplets. Here we consider  examples  of higher derivative  non-unitary   $\mc N=1$   multiplets that 
naturally appear in the context of extended   conformal  supergravities  and  compute 
their superconformal index. We  compare  the  coefficients in the  small $\b$   expansion of the index 
with those  proposed earlier for  unitary multiplets and  suggest  some   modifications that should  apply universally to all types of theories. We also   comment on the structure of subleading terms and 
 the  case of   $\N=4$   conformal supergravity. 
}

\affiliation[a]{Dipartimento di Matematica e Fisica Ennio De Giorgi,\\
Universit\`a del Salento \& INFN, Via Arnesano, 73100 Lecce, 
Italy} 

\affiliation[b]{The Blackett Laboratory, Imperial College, London SW7 2AZ, U.K.}
                       
\emailAdd{matteo.beccaria@le.infn.it} 
\emailAdd{tseytlin@imperial.ac.uk}

\def \N  {{\cal N}}
\def \tI  {\text{I}}

\begin{document}
\date{\currenttime}

\begin{flushright}\small{  Imperial-TP-AT-2018-{04}}\end{flushright}		
\maketitle

\def  \pf {{\phi^{(4)}}}
\def \aa   {{\rm a}}  \def \cc  {{\rm c}}

\section{Introduction}

Unitary representations of  
$\N=1$  superconformal algebra 
 play an important role  in   many aspects  of supersymmetric   quantum field theory.
 Applications  of non-unitary representations are   less  studied. They   appear, in particular, 
   in  extended  conformal supergravities \cite{Kaku:1978nz,Ferrara:1977mv,
   Bergshoeff:1980is,Fradkin:1985am}  written in terms of $\N=1$  multiplets.
Recently, such multiplets were  considered in the  computation  of conformal 
anomalies (in 4 and 6 dimensions)  in the context of  AdS/CFT   
\cite{Beccaria:2014jxa,Beccaria:2014xda, 
Beccaria:2015uta,Beccaria:2015ypa,
Beccaria:2017dmw};  
 similar massless and massive multiplets 
were also  discussed in \cite{Ferrara:2018wqd}.
Non-unitary  superconformal theories  may   
have other interesting applications (see, e.g.,  
 \cite{Buican:2017rya}). 

Given a  CFT  one  may  define   the standard  partition function on $S^1_\beta  \times S^3$ 
(with fermions  treated as  antiperiodic on  $S^1$ of length $\beta$  to have 
  usual    thermodynamic   interpretation).
When  CFT is also a superconformal theory one may
formally  define  also another -- 
"supersymmetric" --  partition function  $Z^\tsusy$ on $S^1_\beta\times S^3$ by 
(i) taking  fermions   to be periodic  on $S^1_\beta$     and (ii) introducing  extra  
 R-symmetry gauge-field couplings in the action to
preserve global supersymmetry on $S^3$ \cite{Dumitrescu:2012ha,Klare:2012gn,Closset:2012ru}.
While having no  thermodynamic  interpretation\foot{We  may  still formally refer  to $\beta$  as  an  inverse "temperature". We shall also assume that the radius of $S^3$ is fixed to be 1.} 
  $Z^\tsusy$   will   instead   be related  to the 
superconformal index $\tI(\b)$   \cite{Romelsberger:2005eg,Kinney:2005ej,Romelsberger:2007ec}. It will  thus  
 be protected by  supersymmetry,  
receiving contributions only from  short multiplets (thus  being computable exactly, {\em e.g.},   using  localization,  see \cite{Rastelli:2016tbz} for a comprehensive review).

In this paper we  shall   study the properties of the superconformal index for  higher-derivative (and  higher spin) 
$\mc N=1$ non-unitary  multiplets. We shall  compute the coefficients   in the 
 small $\b$  expansion of  the index  $\tI(\b)$ 
  and compare   with  their expected expressions 
    in terms of conformal anomaly coefficients   proposed earlier on the basis of  studies 
 of unitary  low-spin examples \cite{DiPietro:2014bca,Ardehali:2015hya}.
We shall  find  that   some modifications of these  expressions   are required in the  
non-unitary cases.\footnote{Our discussion will be restricted to abelian free superconformal theories.
 In presence of a non-trivial
semi-simple gauge group the asymptotic behaviour discussed in \cite{DiPietro:2014bca,Ardehali:2015hya}
 may require corrections when the theory has moduli spaces on the "thermal" cycle 
 \cite{Ardehali:2015bla,Ardehali:2016kza,Aharony:2013kma}, as in the case of the ISS model
 \cite{Intriligator:1994rx,Vartanov:2010xj}, {\em i.e.}  $SU(2)$  $\mc N=1$ SYM
  with a single chiral field in the spin $3/2$ representation. 
  The general reason for such corrections in models with simple gauge groups has been further elucidated
  in \cite{Closset:2017bse,Hwang:2018riu}
  by taking into account contributions from all vacua in the 3d limit.}

We shall start in Section~\ref{sec:review} with a short review of the definition  
of the superconformal index for a 4d  $\mc N=1$ theory  and its relation to the 
supersymmetric partition function on $S^1_\beta\times S^3$. We  shall then 
discuss   what is known
about  the leading coefficients in  their small $\beta$ expansion,   emphasizing 
that
the assumptions   used to derive the  general expressions for the  coefficients    
 were   checked   only in models with  simplest  unitary (chiral and vector)
multiplets. 

In Section~\ref{sec:multiplets}
we  shall  introduce   four basic higher-derivative $\mc N=1$ superconformal
multiplets for which   we shall  later compute the  superconformal index. 
These   non-unitary multiplets are the $\mc N=1$
building blocks of  extended  conformal supergravities.
We shall   discuss  the superfield structure of the multiplets   and  
check the relation between  their chiral and conformal anomalies  as predicted by the superconformal 
invariance.

In Section~\ref{sec:indices} we  shall compute the superconformal index $\tI(\b)$ of these  free non-unitary
multiplets by an explicit "letter"-counting algorithm. 
The multiplets  that  contain higher spin gauge fields
(conformal gravitino and graviton)  require  careful treatment of equation of motion
constraints and Bianchi identities for  the field strengths. In  Section~\ref{sec:zeta} where we present an efficient method to  extract the  small $\beta$ expansion of the  index, including all possible 
subleading terms.

 In Section~\ref{sec:structure} we shall 
 compare the expressions for the  coefficients  in the expansion  of $\tI(\b)$ with those 
 proposed earlier for  unitary multiplets and propose some   modifications that should  apply universally
 to all types of theories. We also   comment on subleading terms and 
 the    case of  finite $\N=4$   conformal supergravity. 
 

In Appendix \ref{app:tensor1} we shall  present the free action of the $[{1\ov 2}]$ tensor multiplet.
In Appendix \ref{app:chiral}  we shall  discuss  the  results for the chiral   gravitational and gauge  anomalies  of the conformal   gravitino and  non-gauge antisymmetric tensor. 
In Appendix \ref{app:CHS}  we shall  review 
 the expression  for the 
 conformal higher spin partition function on $S^1_\beta\times S^3_b$ 
 and work out  its  small $\b$ expansion.
 Appendix \ref{app3d}  will contain a discussion 
 of how to one may compute 
 the constant and $\log \b$  term in the  partition function 
 using the direct expansion and   $\z$-function
  regularization in terms of spectrum of 
 dimensionally reduced  theory on $S^3$. 
In Appendix  \ref{app:squashed}  we  shall repeat the 
computation of the  superconformal  index  and its small $\b$ expansion   for non-unitary multiplets
 in  case of  unequal fugacities 
which is related to   supersymmetric partition function on $S^1_\beta\times S^3_b$
(with $S^3_b$   being a squashed 3-sphere). Finally, in Appendix \ref{app:6d} we 
demonstrate  how 
to perform  similar  analysis  of the index of  six-dimensional theories with $(1,0)$ supersymmetry discussing 
considering the examples of   scalar, tensor and non-unitary higher derivative vector multiplets.

\def \RR  {{\rm R}}

\section{Superconformal index and its small $\beta$  expansion: a  review}
\la{sec:review}

The  superconformal index of an $\mc N=1$  theory on $\mathbb R^4$ is defined 
as  \cite{Romelsberger:2005eg,Kinney:2005ej,Romelsberger:2007ec} 
\be
\la{2.1}
\text{I}(p,q) = \text{Tr}\Big[(-1)^F\,e^{-\mu\,(\Delta-2\,j_2-\frac{3}{2}\,r)}\,
p^{j_1+j_2+\frac{1}{2}\,r}\,q^{-j_1+j_2+\frac{1}{2}\,r}\Big].
\ee
Here, quantum numbers $j_1, j_2, \Delta, r$ label representations of the bosonic compact 
subgroup $SU(2)_{j_1}\times SU(2)_{j_2}\times U(1)_\Delta\times U(1)_r$ of the 
$SU(2,2|1)$ superconformal
group. In particular, $\Delta$ is the conformal dimension and $r$ is 
R-charge.\footnote{ 
Here $j_1$, $j_2$ in (\ref{2.1}) denote the third components of the $SU(2)\times SU(2)$  angular momenta.}
The chemical potentials 
 $p$ and $q$ are free parameters. Due to supersymmetry, the trace 
receives contributions only from states with $\delta \equiv  \Delta-2\,j_2-\frac{3}{2}\,r=0$ and 
 thus $\tI(p,q)$  is 
independent of the third  parameter  $\mu$.
Examples of exact results  for the index obtained by counting or localization can be found in  
\cite{Kinney:2005ej,Romelsberger:2007ec,Bhattacharya:2008zy,
Closset:2013sxa,Benini:2015noa,Honda:2015yha,Benini:2011nc,Razamat:2013opa,
Assel:2014paa,Benini:2016hjo,Closset:2016arn,Nishioka:2014zpa,
Pestun:2007rz,Pestun:2016zxk,DiPietro:2016ond}.

Setting \be p=q\equiv t=e^{-\beta},\la{00}\ee
   one   finds  the special  case  of the
 index   that we shall consider below 
\be
\la{2.4}
\text{I}(\beta) = \text{Tr}\Big[(-1)^F\,e^{-\beta\,(\Delta-\frac{1}{2}\,r)}\Big],\qquad 
\qquad \text{Tr}\equiv\left.\text{Tr}\right|_{\delta=0}.
\ee
This index  that  happens to be directly related 
   to the  supersymmetric partition function $Z^\tsusy(\beta)$ on 
$S^1_\beta\times S^3$  by  \ci{Kim:2012ava,Closset:2013sxa,Assel:2014paa}
 \be
\la{2.2}
Z^\tsusy(\beta) =\ e^{-\beta\,E_\tsusy}
\   \text{I}(\beta) \ . 
\ee
Here $E_\tsusy$ is the "supersymmetric" Casimir energy
 \cite{Kim:2012ava,Assel:2014paa,Lorenzen:2014pna,Assel:2015nca} 
 which can be   expressed in terms of   the conformal anomaly  a  and c coefficients\footnote{For a  conformal
theory on curved space,  the coefficient of the  logarithmic  UV divergence  in the standard partition function is 
\be\notag
\log Z_\infty  
=
\frac{1}{(4\pi)^2} \log\Lambda \int d^4 x\sqrt{g}\,b_4 \ , \qquad  \qquad b_4 = -\text{a}\,R^\star R^\star+\text{c}\,C^2\ . 
\ee
Here   we  ignored    possible  $\nabla^2 R$  term, 
 $C^2$ is the square of the Weyl tensor and $R^\star R^\star = C^2-2R_{\mu\nu}^2+\frac{2}{3}R^2$
is $32\pi^2$ times the Euler  number density.
Note  that in contrast to  the standard Casimir energy \cite{Brown:1977sj,Cappelli:1988vw,Herzog:2013ed}
 the "supersymmetric"  one    may be  viewed as     "scheme-independent"
\cite{Assel:2015nca,Assel:2014tba}  as supersymmetry should prohibit adding extra local counterterms
that   may modify the   the expression for $E_\tsusy$.}
\be
\la{2.3}
E_\tsusy = \frac{4}{27} (\text{a} + 3\,\text{c})\  .
\ee 
%
The small $\beta$ expansion of $Z^\tsusy(\beta)$ takes the  following form 
\be
\la{2.5}
\log Z^\tsusy(\beta) \stackrel{\beta\to 0}{=} C_1\,\frac{\pi^2}{\beta}+C_2+C_3\,\log\beta
+0\cdot\beta
+\mc O(\beta^2),
\ee
where  $C_i$ are theory-dependent numerical coefficients  and 
the absence of the linear in $\beta$ term is due to supersymmetry. 
Then  (\ref{2.2}) implies the following expansion of the 
index \footnote{The fact that the index encodes
 $E_\tsusy$ was first suggested in 
\cite{Kim:2012ava}.
 The relation between the index and $Z^\tsusy$ was later clarified in \cite{Closset:2013sxa}
who showed that the $e^{-\beta\,E_\tsusy}$   factor in (\ref{2.2}) is  a normal-ordering effect like 
 for the standard  Casimir-like contribution. Further  discussion  of 
the universality of the relation (\ref{2.2})  
appeared in  \cite{Assel:2014paa}. 
}
\be
\la{2.6}
\log \text{I}(\beta) \stackrel{\beta\to 0}{=} C_1\,\frac{\pi^2}{\beta}+C_2+C_3\,\log\beta
+E_\tsusy\,\beta
+\mc O(\beta^2).
\ee
A more general specialization than  \rf{00},  depending 
on a 2-parameter family of unequal fugacities $p$ and $q$ in \rf{2.1},
is related \ci{Imamura:2011wg,Closset:2012ru,Aharony:2013dha} 
 to  supersymmetric partition function on $S^1_\b \times S^3_b$    where $S^3_b$
is squashed sphere  and 
   will be discussed in Appendix \ref{app:squashed}.

\

Let us  now review what was claimed  in the  past about each coefficient in the expansion \rf{2.5} or (\ref{2.6}).

\medskip
\noindent {\bf Leading term} $\sim 1/\beta$:

\noindent
It  was  argued   in \cite{DiPietro:2014bca} that the coefficient of the leading Cardy-type \cite{Cardy:1986ie}  term in \rf{2.6}
can   be expressed in terms of the conformal anomaly coefficients as\footnote{Related observations  appeared previously  in 
\cite{Aharony:2013dha,Ardehali:2013gra,Ardehali:2013xya} with further 
developments in \cite{Ardehali:2014zba,Ardehali:2014esa,
Ardehali:2015hya,Shaghoulian:2015kta,Shaghoulian:2015lcn}.
} 
\be
\la{2.7}
C_1 = \frac{16}{3}\,(\text{c}-\text{a}).
\ee
The proof in \cite{DiPietro:2014bca}  was 
based on the expected form of the effective action  of the dimensionally reduced 3d theory   corresponding 
to  the limit of small radius  of $S^1$. It  contains a Chern-Simons term 
$ k \,\int_{S^3} a\wedge F$ 
where $a$ is the Kaluza-Klein graviphoton (mixed component of the metric tensor in 
reduction to 3d) and $F$ is the R-symmetry gauge field strength.
 The coefficient $k\sim C_1$ is then proportional to the R-current  gravitational anomaly. It  
    was  computed by considering the example  of a 4d Weyl  fermion 
 leading to 
\be
\la{2.8}
C_1 = -\frac{1}{3}\,\text{Tr}(\RR) \ , 
\ee
where $\text{Tr}(\RR)$ is the  sum of $r$-charges.   
In general, the  $\mc N=1$ superconformal symmetry relates the  
gravitational R-current anomaly to the trace anomaly
coefficients as   \cite{Anselmi:1997am} 
\be
\la{2.9}
\nabla_\mu R^\mu = -\frac{1}{384\,\pi^2}\text{Tr}(\RR)\,RR^\star+
\frac{1}{16\pi^2}\,\text{Tr}(\RR^3)\,F F^\star
= \frac{\cc-\aa}{24\,\pi^2}\,RR^\star+\frac{5\aa-3\cc}{9\pi^2}FF^\star \ , 
\ee
{\em i.e.} 
\begin{align}
\la{2.10}
&\aa = \tfrac{3}{32}\Big[3\text{Tr}(\RR^{3})-\text{Tr}(\RR)\Big],\qquad \quad \ \,
\cc = \tfrac{1}{32}\Big[9\text{Tr}(\RR^{3})-5\text{Tr}(\RR)\Big],\\
&\text{Tr}(\RR) = 16\,(\text{a}-\text{c})\ , \qquad \qquad\qquad 
 \text{Tr}(\RR^3) = \tfrac{16}{9}\,(5\,\text{a}-3\,\text{c})\ . \la{210} 
\end{align}
Here we use  the somewhat  loose  notation $\text{Tr}(\RR)$ and $\text{Tr}(\RR^3)$    
 for the  gravitational   and  gauge anomaly 
coefficients: they  are literally  the sum of $ r$-charges and their cubes only in the case   of the standard 
 chiral  fermions  but in general  contain also field-dependent  coefficients, {\em i.e.} 
\be
\la{2.11}
\text{Tr}(\RR) \equiv  \sum_i \kappa_{1, i}\,r_i\ ,\qquad \qquad 
\text{Tr}(\RR^3) \equiv  \sum_i \kappa_{3, i}\,r_i^3\ ,
\ee
where $\kappa_1=\kappa_3=1$ for a  left Weyl spinor.

\medskip
\noindent {\bf Constant term}  $\sim\beta^0$:

\noindent
Motivated by the study of  explicit 
examples  of  standard  chiral and vector multiplets,
the  constant $C_2$   in (\ref{2.5}) was   conjectured in \cite{Ardehali:2015hya}
to be  equal to the logarithm of the supersymmetric partition function 
of the dimensionally reduced theory on $S^3$ \footnote{To be precise, 
the Ansatz in \cite{Ardehali:2015hya} is slightly different  by a multiple
of $\log(2\pi)$ because the logarithmic term 
in (\ref{2.5})  or (\ref{2.6}) is written as 
$\log\frac{\beta}{2\pi}$. We claim that the identification (\ref{2.111}) is correct if 
the logarithmic term is simply $\log\beta$, see Appendix  \ref{app3d}.
}
\be 
\la{2.111}
C_2= {\rm k}\equiv \log Z^\tsusy_{S^3} \ . \ee

\medskip
\noindent {\bf Logarithmic term} $\sim\log\beta$:

\noindent
The  coefficient $C_3$  of the logarithmic term in (\ref{2.5}) was  conjectured to be \cite{Ardehali:2015hya}
\be
\la{2.12}
C_3=-4\,(2\text{a}-\text{c}),
\ee
again  motivated   by  the  examples of   chiral and  vector multiplets.\foot{Let us note  that the combination 
$2\,\text{a}-\text{c}$ 
plays a  special role in the analysis of the $\mc N=2$ models in \cite{Shapere:2008zf,Buican:2014qla}
where it is essentially the sum of  dimensions of operators parametrizing the Coulomb branch.}

\medskip
\noindent {\bf Linear term} $\sim\beta$:

\noindent
As already mentioned, the  coefficient of the  linear term in \rf{2.6} 
is the "supersymmetric" analog of Casimir energy.\footnote{The effective Hamiltonian appearing in \rf{2.4} is 
$H_\tsusy=\Delta-\frac{1}{2}\RR$. 
Normal ordering 
:${H_\tsusy}$: $ = H_\tsusy-\langle H_\tsusy\rangle = H_\tsusy-E_\tsusy$ is  implicitly  understood,
and  is essentially the   reason why $Z^\tsusy$ is not equal to the index in \rf{2.2}
(see 
 \cite{Closset:2013sxa} for details). 
This is  a general feature of the relation between the QFT partition function and 
the thermodynamic  partition function Tr$( e^{-\beta H})$   \cite{Allen:1986qi}.
}
The relation (\ref{2.2})  may  be studied in the  low temperature $\beta\to \infty$
or high temperature $\beta\to 0$ limit.  For $\beta\to \infty$ the leading contribution comes only from the 
energy  exponential, {\em i.e.} 
\be
\la{2.13}
Z^\tsusy \stackrel{\beta\to\infty}{\sim} e^{-\beta\,E_\tsusy},\qquad \qquad \text{I}(\beta) 
\stackrel{\beta\to\infty}{\sim} 1\ , 
\ee
where the  asymptotics of the index corresponds to  the vacuum contribution.
Equivalently,  
$E_\tsusy = -\lim_{\beta\to \infty}\frac{d}{d\beta}\log Z^\tsusy$.
On the other hand, in the  $\beta\to 0$ limit  the partition function   (and the index) 
is governed by the high-energy part of the spectrum. This leads to the  
 singular  Cardy-type   term in \rf{2.5}.\footnote{For comparison, let us mention   what  happens 
in  the case of the standard partition function of a non-supersymmetric  
CFT. For example, for a free 
conformal scalar  the derivative $\mc E(\beta) \equiv  -\partial_\beta \log\text{Tr}(e^{-\beta \Delta})$ obeys
$\mc E(\beta) = \frac{\pi ^4}{15 \beta ^4}-\frac{1}{240}+\big(\frac{2\pi}{\beta}\big)^4\,\mc E(\frac{2\pi}{\beta})$.  For $\beta\to 0$ we get 
\ $\mc E(\beta) \stackrel{\beta\to 0}{=} \frac{\pi ^4}{15 \beta ^4}-\frac{1}{240}+\dots$
(up to exponentially suppressed terms). This  implies
$\log\text{Tr}(e^{-\beta \Delta}) \stackrel{\beta\to 0}{=} \frac{C}{\beta^3}+E_\text{cas}\,\beta + \dots.$,
where  $E_\text{cas}$ is the standard Casimir energy $\langle H\rangle$=$\langle\Delta\rangle$.
}
The  relation  \rf{2.3} of  $E_\tsusy$   to  the conformal anomaly coefficients  was  demonstrated 
explicitly in the case of  the chiral and vector multiplets \ci{Closset:2013sxa,Assel:2014paa}.  
In addition, a
 general derivation was  proposed  in  \cite{Bobev:2015kza} based on 
a representation of $E_\tsusy$ in terms of the anomaly polynomial
and assuming the  standard   relations \rf{2.10}  between the anomaly coefficients.

\medskip
\noindent {\bf Higher order terms $\sim \beta^n$:}

\noindent
It was claimed in  \cite{Ardehali:2015hya} that the $\mc O(\beta^2)$ corrections are 
 exponentially suppressed, {\em i.e.} schematically go as $e^{-1/\b}$ for $\beta\to 0$.
 Let us recall,  as an analogy,   that 
absence of $\beta^n$ ($n\ge 2$) corrections for in the  case of the logarithm of the thermodynamic 
partition  function $Z(\b)$  was observed  in the past for  the standard conformal fields
\cite{Kutasov:2000td} where this  follows  from a simple modular transformation property  of $Z(\b)$.
 It is unclear if   similar modular properties play a role in the supersymmetric context.

\


The above discussion   leaves  several open questions. 
One is  whether  the  prediction  \rf{2.8}   and thus (\ref{2.7})  for the leading-order coefficient $C_1$ 
 is completely  universal, {\em i.e.}   holds for   general  $\mc N=1$ superconformal theories, including  also non-unitary 
and  higher spin  multiplets.   The argument   in \cite{DiPietro:2014bca}    relying on  Chern-Simons term in the 
 reduced 3d effective theory  appears  to be specific to  case of standard Weyl fermions. 
 As we shall see   below,  the relation   \rf{2.8}   indeed requires  a    generalization in the non-unitary case. 

The second question is about the    expression \rf{2.12}  for  coefficient  $C_3$ of the  $\log \b$ term
which was  checked only  for   standard  multiplets. 
We shall find  that   \rf{2.12}   needs 
  a modification in the case of  higher spin   superconformal multiplets. 
  We shall   propose an alternative universal   expression   for  the 
  $\log \b$ term  in terms of the integer numbers of   conformal Killing tensors 
  associated with each   conformal gauge  field  in the multiplet.  

Finally,   it is not clear  a priori if  the  non-unitary  multiplets   will also   have  exponentially decaying 
  corrections  in  their  index expansion, {\em i.e.}   if all  power $\b^2,\b^3, ...$ corrections  in \rf{2.6}  
  will be absent.  
  Indeed, we shall find that such power corrections will survive for multiplets containing higher spin ($s>1$)
  conformal spins.  
  

To address  these questions,  below  we  shall  consider  several non-unitary (higher derivative  and  higher spin)
superconformal  multiplets  that appear in the context 
 of extended conformal supergravities. For each of  these  free multiplets
we will explicitly compute the index  \rf{2.4},  obtain the coefficients in  its  small $\beta$ expansion
and compare them  with their expected values      based on relations found  earlier in the 
   studies of  unitary multiplets. 


\section{Higher derivative $\N=1$     superconformal multiplets}
\la{sec:multiplets}

Here    we will describe  the  content of the  four   basic higher-derivative $\mc N=1$ 
superconformal multiplets  for which  we    will  later compute the  index \rf{2.4}. 
They naturally appear   in the decomposition of $\N \leq 4$ conformal supergravities in terms of $\N=1$ multiplets.

\subsection{ $\mc N=1$  multiplet content of  extended  conformal supergravities}

Let us start with reviewing  the  field content  of  conformal   supergravities (for details see  \cite{Bergshoeff:1980is,Fradkin:1985am}). 
Expanded near flat-space vacuum  they are   given  by a collection of 
the following  free conformal fields: 

%
\begin{tabbing}

$\phi$:  \qquad \= standard 2-derivative real scalar,  $L\sim \phi \Box \phi$\\

$\pf $:  \> 4-derivative real scalar,   $L \sim \pf \Box^2\pf $\\

$\psi$:  \> standard  Weyl fermion,   $L\sim \bar \psi  \slashed{\partial}  \psi$\\

$\psi^{(3)} $:   \>  3-derivative  Weyl fermion,   $L\sim \bar \psi^{(3)}  \slashed{\partial}^3  \psi^{(3)}$\\

$V_\mu$:   \>  standard   gauge  vector,   $L \sim F_{\m\n}  F^{\m\n} $\\

$T_{\m\n}$:  \>  non-gauge  real antisymmetric tensor,  $L\sim \partial^\mu T_{\mu\nu}^+\partial_\lambda T^{- \lambda\nu}$, \ \ 
$T^\pm = T \pm T^*$\\


 $\Psi_\mu$: \> conformal gravitino,  
$L \sim \overline\Psi_\mu\slashed{\partial}^3\Psi^\mu$\\

$h_{\mu\nu}$: \>     conformal (Weyl)  graviton,     $L\sim  h  \Box^2 h$ 
\end{tabbing}

\noindent
  The $\mc N=1$  multiplet content of  extended conformal supergravities (CSG) is \ci{Fradkin:1985am}
\begin{align}
\la{3.1}
\mc N=1\ \text{CSG} &= [2],\notag\\
\mc N=2\ \text{CSG} &= [2]+[\tfrac{3}{2}]+[1],\notag\\
\mc N=3\ \text{CSG} &= [2]+2\,[\tfrac{3}{2}]+4\,[1]+[\tfrac{1}{2}]+2\,[0],\notag\\
\mc N=4\ \text{CSG} &= [2]+3\,[\tfrac{3}{2}]+8\,[1]+3\,[\tfrac{1}{2}]+6\,[0]+[0'] \ .
\end{align}
Here $[0] = ( 2 \phi, \psi)$  and $[1]= (V_\mu, \psi)$ are the standard unitary  scalar and vector multiplets while 
$[2]$,  $[\tfrac{3}{2}]$, $[\tfrac{1}{2}]$   and $[0']$  are the four   non-unitary multiplets containing higher derivative fields: 
 \begin{align} \la{33.1} 
& \left[0'\right]=(2\,\pf , \psi^{(3)}, 2\,\phi)\ , \qquad \qquad 
\left[\tfrac{1}{2}\right]=(\psi, 2\, \phi, T_{\mu\nu}, \psi^{(3)})\ , \qquad\no \\
& \left[\tfrac{3}{2}\right]=(\Psi_\mu, 2\,V_\mu, T_{\mu\nu}, \psi) \ , \qquad \qquad 
\left[2\right]=(h_{\mu\nu}, \Psi_\mu, V_\mu) \ .
\end{align} 
The maximal $\mc N=4$ supergravity  has  local $SU(4)$ R-symmetry
under which the fields  transform in  various representations. 
 Decomposing $SU(4)\to SU(3)\times U(1)$ allows one  to identify the
$U(1)$ with the $\mc N=1$ R-symmetry  corresponding to  the $\mc N=1$ multiplets
and thus fix  the  $r$-charges of the component fields. 
In particular, the conformal graviton is a singlet  so has $r=0$ 
while for other fields   one finds 
\begin{align}
\la{3.2}
\Psi_\mu :\quad  & \bm{4} = \mathop{\bm{1}_1}_{[2]}+\mathop{\bm{3}_{-1/3}}_{3\,[\frac{3}{2}]}
, \qquad \qquad 
T_{\m\n}  :\quad  \bm{6} = \mathop{\bm{3}_\frac{2}{3}}_{3\,[\frac{3}{2}]}
+\mathop{\overline{\bm{3}}_{-\frac{2}{3}}}_{3\,[\frac{1}{2}]},\notag \\
V_\mu :\quad & \bm{15} = \mathop{\bm{1}_0}_{[2]}
+\underbrace{\bm{3}_{-4/3}+\overline{\bm{3}}_{4/3}}_{3\,[\frac{3}{2}]}
+\mathop{\bm{8}_0}_{[1]}, \notag \\
\pf  :\quad & \mathop{2\times\bm{1}_0}_{[0']}, \qquad \qquad \qquad \quad\ \,
\psi^{(3)} :\quad  \overline{\bm{4}}=\mathop{\bm{1}_{-1}}_{[0']}
+\mathop{\overline{\bm{3}}_{1/3}}_{3\,[\frac{1}{2}]}, \notag \\
\phi :\quad & 2\times \overline{\bm{10}} = \mathop{2\times\bm{1}_{-2}}_{[0']}
+\mathop{2\times\overline{\bm{3}}_{-2/3}}_{3\,[\frac{1}{2}]}
\mathop{+2\times\overline{\bm{6}}_{2/3}}_{6\,[0]}, \notag \\
\psi:\quad & \bm{20}=\mathop{\bm{3}_{-1/3}}_{3\,[\frac{3}{2}]}+
\mathop{\overline{\bm{3}}_{-5/3}}_{3\,[\frac{1}{2}]}
+\mathop{\overline{\bm{6}}_{-1/3}}_{6[0]}
+\mathop{\bm{8}_1}_{8\,[1]}.
\end{align}
Here  we indicated  how the  $SU(4)$ representations  of each field in $\N=4$ theory is   decomposed, resulting $\N=1$ $R$-charges
and   the $\N=1$  multiplets    each  field belongs to. 
For example,  the 4   gravitinos   of the  $\mc N=4$  supergravity 
 split  into the one belonging to  $[2]$ and having  $r=1$, and three from the three $[\frac{3}{2}$] multiplets with $r=-\frac{1}{3}$. 
The  resulting  R-charge values are in  agreement with the representation
theory of $\mc N=1$ superconformal algebra discussed  below. 

For future reference   in Table \ref{tab:ac}   we list  the conformal anomaly coefficients  of the  individual fields
(see \ci{Fradkin:1981jc,Fradkin:1982xc,Fradkin:1985am}).
\begin{table}
\setlength\extrarowheight{8pt}
\begin{center}
\begin{tabular}{lcccccccc}
\toprule
& $\phi$ & $\pf $ & $\psi$ & $\psi^{(3)}$ & 
$T_{\mu\nu}$ & $V_{\mu}$ & $\Psi_{\mu}$ & $h_{\mu\nu}$ \\
\midrule
a & $\frac{1}{360}$ &$ -\frac{7}{90}$ &$ \frac{11}{720}$
&$ -\frac{3}{80}$ &$ -\frac{19}{60}$ &$ \frac{31}{180}$
&$ -\frac{137}{90}$ &$ \frac{87}{20}$\\
c & $\frac{1}{120}$ &$ -\frac{1}{15}$ &$ \frac{1}{40} $
&$ -\frac{1}{120}$ &$ \frac{1}{20}$ &$ \frac{1}{10}$
&$ -\frac{149}{60}$ &$ \frac{199}{30}$\\
\bottomrule
\end{tabular}
\end{center}
\caption{Conformal anomaly coefficients of  fields appearing in $\mc N=1$ superconformal
multiplets.}
\la{tab:ac} 
\end{table}
In Table \ref{tab:multiplets}   we present the resulting values of a  and c for the  basic  unitary and non-unitary  multiplets
introduced above.  We  also  give   particular  combinations  of a  and c   corresponding to  the expected  
values of 
 (i)  the coefficients $\text{Tr}(\RR)$ and 
$\text{Tr}(\RR^3)$  of the   gravitational  and R-symmetry  chiral  anomalies  computed  according to \rf{210}
(ii)   $E_\tsusy$  as given by \rf{2.3},  and  (iii)  the coefficient $C_3$  as defined  in \rf{2.12}. 
\begin{table}
\setlength\extrarowheight{5pt}
\begin{center}
\begin{tabular}{clcccccc}
\toprule
& & a & c & $\text{Tr}(\RR)$ & $\text{Tr}(\RR^3)$ & $E_\tsusy$
& $C_3$ \\
\midrule
$\left[0\right]$ & $(2\phi,\psi)$  
& $\frac{1}{48}$ & $\frac{1}{24}$ & $-\frac{1}{3}$ & $-\frac{1}{27}$ & $\frac{7}{324}$ & 0\\
$\left[1\right]$ & 
$(V_\mu,\psi)$  & $\frac{3}{16}$ & $\frac{1}{8}$ & 1 & 1 & $\frac{1}{12}$ & -1\\
\midrule
$\left[0'\right]$ &  $(2\,\pf , \psi^{(3)}, 2\,\phi)$  & $-\frac{3}{16}$ 
& $-\frac{1}{8}$ & -1 & -1 & $-\frac{1}{12}$& 1 \\
$\left[\tfrac{1}{2}\right]$ &  $(\psi, 2\phi, T_{\mu\nu}, \psi^{(3)})$  & $-\frac{1}{3}$ 
& $\frac{1}{12}$ & $-\frac{20}{3}$ & $-\frac{92}{27}$ 
& $-\frac{1}{81}$ & 3\\
$\left[\tfrac{3}{2}\right]$ &  $(\Psi_\mu, 2\,V_\mu, T_{\mu\nu}, \psi) $ 
& $-\frac{71}{48}$ & $-\frac{53}{24}$ & $\frac{35}{3}$ & $-\frac{37}{27}$
& $-\frac{389}{324}$ & 3 \\
$\left[2\right]$ &  $(h_{\mu\nu}, \Psi_\mu, V_\mu)$  & 3 & $\frac{17}{4}$ & -20 & 4 
& $\frac{7}{3}$ & -7 \\
\bottomrule
\end{tabular}
\end{center}
\caption{The values of the conformal anomaly coefficients  and their  combinations 
$\text{Tr}(\RR) = 16\,(\text{a}-\text{c})$, 
$\text{Tr}(\RR^3) = \tfrac{16}{9}\,(5\,\text{a}-3\,\text{c})$, $E_\text{\tiny SUSY}=\frac{4}{27}(\text{a}+3\text{c})$, and $C_3=-4(2\text{a}-\text{c})$ 
for the 2  unitary and 4 non-unitary   superconformal $\mc N=1$ multiplets. 
}
\la{tab:multiplets} 
\end{table}
It is interesting to  observe  \ci{Fradkin:1985am}  that  the values   for the $[1]$ and $[0']$    multiplets are exactly 
 opposite to each  other.  Note also that  the combinations  
 $3\,\text{Tr}(\RR)$    and        $C_3$  are  always  integer.

\subsection{Structure  of the $\mc N=1$ multiplets }

Let us now   discuss  the  $\mc N=1$ superfield    description   of the  above multiplets
that allows  one to independently  fix  the  R-charges   of the  individual fields  which will be in agreement 
 the  R-charge assignment  in (\ref{3.2})  following  from the   $\mc N=4$   conformal supergravity. 
 
As a consistency check, we will then   be able to show that the resulting  chiral anomaly coefficients 
 $\text{Tr}(\RR)$ and $\text{Tr}(\RR^3)$ computed   directly from  \rf{2.11}  will be in agreement with the values in Table 
  \ref{tab:multiplets}   found assuming  the supersymmetry-implied relations \rf{210}. 
  
 To this end we will   need, in addition to the values of R-charges $r_i$,
  also  the field-dependent   chiral  gravitational and gauge 
 anomaly 
 coefficients $\kappa_1$ and $\kappa_3$ in (\ref{2.11}). 
 Their values 
  for the fields   contributing to the chiral  anomalies  -- 
   Weyl fermions $\psi,\psi^{(3)} \sim  (\frac{1}{2},0)$, Weyl  conformal gravitino  $\Psi_\mu \sim  (1,\frac{1}{2})$
  and self-dual tensor   $T^+_{\m\n}\sim(1,0)$ are   given by (see \cite{Christensen:1978gi,Romer:1985yg,Frampton:1985mb,Carrasco:2013ypa})
\be
\la{3.4}
\def\arraystretch{1.3}
\begin{array}{ccccc}
\toprule
& \psi & \psi^{(3)} & T^+_{\m\n} & \Psi_\mu \\
\midrule
\kappa_1\ \  & 1 & 1 & 8 & -20 \\
\kappa_3\ \  & 1 & 1 & -4 & 4  \\
\bottomrule
\end{array}
\ee
The chiral anomaly  does not depend on extra derivatives in the kinetic term and thus is the  same for $\psi$ and $\psi^{(3)}$. 
The Lorentz  index of the gravitino is inert   under R-symmetry  and thus its  gauge anomaly is 4 times that of the  Weyl spinor  (cf. \ci{Nielsen:1984es}). 
The non-trivial cases  of the antisymmetric non-gauge  tensor and conformal gravitino  are further
reviewed   in Appendix \ref{app:chiral}.

\subsubsection{Unitary  scalar  $[0]$ and  vector $[1]$  multiplets}

The $[0]$ chiral multiplet     containing  one complex scalar 
and one Weyl fermion corresponds to 
 a chiral superfield $\Phi =  \phi + \theta  \psi + \theta^2 \varphi$. Omitting   the auxiliary field $\varphi$ and 
 using that $\theta^\alpha $   has R-charge 1   we then find  the  following dimensions $\Delta$   and R-charges 
\be
\la{3.5}
\def\arraystretch{1.3}
\begin{array}{cccc}
\toprule
& \Delta & (j_1, j_2) & r  \\
\midrule
\phi  & 1 & (0,0) & r \\
\psi_\alpha & \frac{3}{2} & (\frac{1}{2},0) & r-1\\
\bottomrule
\end{array}\qquad\qquad
\begin{array}{cccc}
\toprule
 & \Delta & (j_1, j_2) & r  \\
\midrule
\overline\phi & 1 & (0,0) & -r\\
\overline\psi_{\dot\alpha} & \frac{3}{2} & (0,\frac{1}{2}) & -r+1\\
\bottomrule
\end{array}
\ee
The $\mc N=1$ superconformal algebra requires that  for the 
superconformal primary   or the lowest chiral superfield   component 
(here the scalar field) one should have  $r=\frac{2}{3}\Delta $   (see, {\em e.g.},  
\cite{Dolan:2008qi}).
This fixes  $r=\frac{2}{3}$,
consistently with the $SU(4)$ decomposition in (\ref{3.2}). 
The resulting  chiral anomaly coefficients  in \rf{2.11},\rf{3.4}  are then 
\begin{align}
\la{3.6}
\text{Tr}(\RR) = \underbrace{r-1}_{\psi} = -\tfrac{1}{3}, \qquad\qquad 
\text{Tr}(\RR^{3}) = \underbrace{(r-1)^{3}}_{\psi}= -\tfrac{1}{27} \ ,  
\end{align}
  in  agreement with the  values  in Table \ref{tab:multiplets}. 
Similar  agreement will  be found  for all other multiplets  discussed below.

The  vector multiplet   $[1]$  is related to    the  
  chiral   spinor   field  strength superfield $W_\a = \psi_\a  + \theta^\b F^+_{\a\b} + ...$
 so that  we find 
\be
\la{3.7}
\def\arraystretch{1.3}
\begin{array}{cccc}
\toprule
& \Delta & (j_1, j_2) & r  \\
\midrule
\psi_\alpha  & \frac{3}{2} & (\frac{1}{2},0) & r \\
F^+_{\alpha\beta} & 2 & (1,0) & r-1\\
\bottomrule
\end{array}\qquad\qquad
\begin{array}{cccc}
\toprule
 & \Delta & (j_1, j_2) & r  \\
\midrule
\overline\psi_{\dot\alpha} & \frac{3}{2} & (0,\frac{1}{2}) & -r\\
F^-_{\dot\alpha\dot\beta} & 2 & (0,1) & -r+1\\
\bottomrule
\end{array}
\ee
The symmetric tensors $F^+_{\alpha\beta}$ and $F^-_{\dot\alpha\dot\beta}$ 
 are the (anti) self-dual parts of the Maxwell field strength in   spinor notation. 
Here the lowest  component  is $\psi_\a$ so that   $r = \frac{2}{3}\Delta_\psi = 1$.
This gives, in agreement  
  with  Table \ref{tab:multiplets},  
\begin{align}
\la{3.8}
\text{Tr}(\RR) = \underbrace{r}_{\psi} = 1, \qquad\qquad 
\text{Tr}(\RR^{3}) = \underbrace{r^{3}}_{\psi}= 1.
\end{align}

\subsubsection{Higher derivative scalar multiplet $[0']$}

A general discussion of $\mc N=1$
non-unitary multiplets
can be found in  \cite{Li:2014gpa} (at the level of states), and 
in \cite{Kugo:1983mv} (at the level of fields). 
We need to   embed  the  $[0']$ multiplet in \rf{33.1}  into a   chiral superfield  
with the  superconformal primary  being  the 4-derivative scalar $\pf $ of dimension 0.
Other fields   are then obtained by  applications of the supersymmetry generator $Q_\alpha\sim (\frac{1}{2},0)$
leading to 
\be
\la{3.9}
\def\arraystretch{1.3}
\begin{array}{cccc}
\toprule
& \Delta & (j_1, j_2) & r  \\
\midrule
\pf  & 0 & (0,0) & r \\
\psi^{(3)}_\alpha & \frac{1}{2} & (\frac{1}{2},0) & r-1\\
\phi & 1 & (0,0) & r-2\\
\bottomrule
\end{array}\qquad\qquad
\begin{array}{cccc}
\toprule
& \Delta & (j_1, j_2) & r  \\
\midrule
\overline{\phi}^{(2)} & 0 & (0,0) & -r\\
\overline\psi^{(3)}_{\dot\alpha} & \frac{1}{2} & (0,\frac{1}{2}) & -r+1\\
\bar\phi & 1 & (0,0) & -r+2\\
\bottomrule
\end{array}
\ee
This  multiplet may be viewed as 
resulting from the application of extra $\Box$ to the   standard chiral  multiplet, {\em i.e.} 
$(\phi, \psi, \varphi)\to (\pf , \psi^{(3)}, \phi)$,  where, in particular,  the auxiliary  field $\vp$ 
becomes a  (complex) dynamical scalar. 
As this multiplet  is non-unitary, it is not  a priori  obvious how to fix   the value of $r$. 
Nevertheless, from the analysis of  \cite{Kugo:1983mv} (see also  \cite{Yamada:2016tca})
the  vanishing scaling dimension 
$\Delta=0$  should imply  $r=0$, {\em i.e.} vanishing chiral weight. 
In practice,  this is still consistent   with  the rule $r=\frac{2}{3}\Delta_{\pf }=0$. 
From the point of view
of $\mc N=4$   supergravity (cf. \rf{3.2}) 
this is  also  consistent with the higher derivative scalar     being an $SU(4)$ singlet.
The direct evaluation of $\text{Tr}(\RR)$ and $\text{Tr}(\RR^{3})$    then gives 
\begin{align}
\la{3.10}
\text{Tr}(\RR) = \underbrace{r-1}_{\psi^{(3)}} = -1, \qquad\qquad 
\text{Tr}(\RR^{3}) = \underbrace{(r-1)^{3}}_{\psi^{(3)}}= -1 \ , 
\end{align}
in agreement with the values in   Table \ref{tab:multiplets}.

\subsubsection{Tensor multiplet $[\frac{1}{2}]$}

The superfield embedding  of this multiplet  can be found, e.g., in   \cite{Louis:2007nd}.\footnote{In   \cite{Louis:2007nd}
the antisymmetric tensor component   is the standard  gauge-invariant one   but this  is not relevant for the purpose 
of fixing R-charges we are concerned with here.} 
The  field content  in \rf{33.1}  may be organized into a chiral superfield, see  Appendix \ref{app:tensor1}.
Its   lowest component is $\psi^{(3)}_\alpha$   and other components 
are built by acting with $Q_\alpha$  decreasing $r$ by one (or, in conjugate   case,  with 
 $\overline Q_{\dot\alpha}$ increasing $r$ by one). As a result, we find 
\be
\la{3.11}
\def\arraystretch{1.3}
\begin{array}{cccc}
\toprule
& \Delta & (j_1, j_2) & r  \\
\midrule
\psi^{(3)}_\alpha & \frac{1}{2} & (\frac{1}{2},0) & r\\
\phi & 1 & (0,0) & r-1\\
T^+_{\alpha\beta} & 1 & (1,0) & r-1\\
\psi_\alpha & \frac{3}{2} & (\frac{1}{2},0) & r-2\\
\bottomrule
\end{array}\qquad\qquad
\begin{array}{cccc}
\toprule
& \Delta & (j_1, j_2) & r  \\
\midrule
\overline\psi^{(3)}_{\dot\alpha} & \frac{1}{2} & (0,\frac{1}{2}) & -r\\
\overline\phi & 1 & (0,0) & -r+1\\
T^-_{\dot\alpha\dot\beta} & 1 & (0,1) & -r+1\\
\overline\psi_{\dot\alpha} & \frac{3}{2} & (0,\frac{1}{2}) & -r+2\\
\bottomrule
\end{array}
\ee
The R-charge is again determined by $r=\frac{2}{3}\Delta_{\psi^{(3)}} = \frac{1}{3}$.
The direct computation of the chiral anomaly coefficients  $\text{Tr}(\RR)$ and $\text{Tr}(\RR^3)$ 
based on \rf{2.11},\rf{3.4}  involves   summing up contributions from all  the  fields  but the scalars 
\begin{align}
\la{3.12}
\text{Tr}(\RR) &= \underbrace{r}_{\psi^{(3)}}\underbrace{+8\,(r-1)}_{T^+}
\underbrace{+r-2}_{\psi} = \tfrac{1}{3}+8\times(-\tfrac{2}{3})-\tfrac{5}{3} = -\tfrac{20}{3}, \notag \\
\text{Tr}(\RR^{3}) &= \underbrace{r^{3}}_{\psi^{(3)}}\underbrace{-4\,(r-1)^{3}}_{T^+}
\underbrace{+(r-2)^{3}}_{\psi} = (\tfrac{1}{3})^{3}-4\times(-\tfrac{2}{3})^{3}
+(-\tfrac{5}{3})^{3} = -\tfrac{92}{27}.
\end{align}

\def \VV {{\rm V}} \def \FF {{\rm F}}

\subsubsection{Conformal gravitino multiplet $[\frac{3}{2}]$}

A  review of the conformal gravitino supermultiplet can be found in Appendix  C of \cite{Becker:2017zwe}.
It   was  also discussed recently in \cite{Kuzenko:2017ujh}  in the context of 
higher spin generalizations. 
In general, one can consider a superconformal multiplet associated with an  integer superspin $s$
and  described in terms of an unconstrained  superfield 
\be
\la{3.13}
\Psi_{\alpha(s)\,\dot\alpha(s-1)}\equiv\Psi_{\alpha_1\dots\alpha_s\dot\alpha_1\dots\dot\alpha_{s-1}},
\qquad
\overline\Psi_{\alpha(s-1)\,\dot\alpha(s)}\equiv\overline\Psi_{\alpha_1\dots\alpha_{s-1}
\dot\alpha_1\dots\dot\alpha_{s}},
\ee
where $\alpha(s)$ denotes a set of $s$ symmetrized indices. 
The gauge freedom is 
\begin{align} 
\la{3.14}
&s >1: \ \ \ \   \delta\Psi_{\alpha_1\dots\alpha_s\dot\alpha_1\dots\dot\alpha_{s-1}}
= D_{(\alpha_1}\overline\Lambda_{\alpha_2\dots\alpha_s)\dot\alpha_1\dots\dot\alpha_{s-1}}
+\overline D_{(\dot\alpha_1}\zeta_{\alpha_1\dots\alpha_s\dot\alpha_2\dots\dot\alpha_{s-1})},
  \\
&s=1: \ \ \ \ \delta\Psi_\alpha = D_\alpha\overline\Lambda+\zeta_\alpha \ ,  \la{3.15}
\end{align} 
with unconstrained gauge parameters $\overline \Lambda_{\alpha(s-1)\dot\alpha(s-1)}$
and $\zeta_{\alpha(s)\dot\alpha(s-2)}$. 
The superfield $\Psi_{\alpha(s)\,\dot\alpha(s-1)}$ has superconformal weights 
$(q, \overline q)=(-\frac{s}{2},\frac{1-s}{2})$  
  \cite{Kuzenko:2017ujh}, so that 
 its dimension and R-charge are $\Delta = q+\overline q = \frac{1}{2}-s$ and 
$r=\frac{2}{3}(q-\overline q)
= -\frac{1}{3}$. 
One may choose a Wess-Zumino gauge  where 
\begin{align}
\la{3.16}
\Psi_{\alpha_1\dots\alpha_s\dot\alpha_1\dots\dot\alpha_{s-1}}(\theta,\overline\theta) &= 
\theta^\beta\overline\theta^{\dot\beta}\psi_{(\beta\alpha_1\dots\alpha_s)(\dot\beta\dot\alpha_1\dots
\dot\alpha_{s-1})}+\overline\theta^2\theta^\beta T_{(\beta\alpha_1\dots\alpha_s)
\dot\alpha_1\dots\dot\alpha_{s-1}}\notag \\
&\ \ \   -\theta^2\overline\theta^{\dot\beta}\
\VV_{\alpha_1\dots\alpha_s(\dot\beta\dot\alpha_1\dots\dot\alpha_{s-1})}
+\theta^2\overline\theta^2\psi_{\alpha_1\dots\alpha_s\dot\alpha_1\dots\dot\alpha_{s-1}},
\end{align}
with {complex} bosonic fields $\VV_{\alpha_1\dots\alpha_s(\dot\beta\dot\alpha_1\dots\dot\alpha_{s-1})} = (V+i\, V')_{\alpha_1\dots\alpha_s(\dot\beta\dot\alpha_1\dots\dot\alpha_{s-1})}$ and $T_{\alpha(s+1)\dot\alpha(s-1)}$.
Specialization to  our case of interest $s=1$   gives
\begin{align}
\la{3.17}
\Psi_{\alpha}(\theta,\overline\theta) &= 
\theta^\beta\overline\theta^{\dot\beta}\psi_{(\alpha\beta)\dot\beta}+\overline\theta^2\theta^\beta 
T_{(\alpha\beta)}
 -\theta^2\overline\theta^{\dot\beta}\
\VV_{\alpha\dot\beta}
+\theta^2\overline\theta^2\psi_{\alpha} \ , 
\end{align}
with the same   field content as in \rf{33.1}. 
For $s>1$ the fields $\VV_{\alpha(s)\dot\alpha(s)}$ and $T_{\alpha(s+1)\dot\alpha(s-1)}$ have 
 residual gauge invariances.  In the special   case of $s=1$ the field  $T_{(\alpha\beta)}\equiv T^+_{\a\b}$
is a non-gauge one   \cite{Kuzenko:2017ujh}.
From (\ref{3.17}) 
we find   (here $\Psi_{(\alpha\beta)\dot\beta}$ 
 is the gravitino in spinor notation) 
\be
\la{3.18}
\def\arraystretch{1.3}
\begin{array}{cccc}
\toprule
 & \Delta & (j_1, j_2) & r  \\
\midrule
\Psi_{(\alpha\beta)\dot\beta} & \frac{1}{2} & (1,\frac{1}{2}) & r \\
\VV_{\alpha\dot\alpha} & 1 & (\frac{1}{2},\frac{1}{2}) & r-1\\
T^+_{\alpha\beta} & 1 & (1,0) & r+1\\
\psi_\alpha & \frac{3}{2} & (\frac{1}{2},0) & r\\
\bottomrule
\end{array}\qquad\qquad
\begin{array}{cccc}
\toprule
& \Delta & (j_1, j_2) & r  \\
\midrule
\overline\Psi_{(\dot\alpha\dot\beta)\beta} & \frac{1}{2} & (\frac{1}{2},1) & -r \\
\overline \VV_{\alpha\dot\alpha} & 1 & (\frac{1}{2},\frac{1}{2}) & -r+1\\
T^-_{\dot\alpha\dot\beta} & 1 & (0,1) & -r-1\\
\overline\psi_{\dot\alpha} & \frac{3}{2} & (0,\frac{1}{2}) & -r\\
\bottomrule
\end{array}
\ee
According to the above general discussion  here we should    have  $r=-\frac{1}{3}$.

It   is useful   to  consider also  an alternative  and more transparent  description 
 of the $[{3\ov 2}]$  multiplet   in terms of  the  gauge-invariant  chiral   superfield  
 (see  \ci{Siegel:1980bp}) 
\be
\la{3.19}
W_{\alpha\beta} = T^+_{\alpha\beta}+\theta^{\gamma}\,(
\Psi_{\alpha\beta\gamma}+\eps_{\gamma (\alpha}\psi_{\beta)})+
\theta^{2}\,\FF_{\alpha\beta} \ , 
\ee
where  $\Psi_{\alpha\beta\gamma}$ is  the gravitino field strength ({\em i.e.}  the self-dual part of 
$\del_{[\m} \Psi_{\nu]}$) and $\FF$ is the field strength of the complex vector. 
Here the  dimensions  of the components are $\Delta=1,\frac{3}{2},2$ and the R-charges are  $r+1, r, r-1$
(here we set  $r_{_T}\equiv  r+1$   to match the notation in \rf{3.18}). As this is a chiral  superfield, its  
  lowest  component   should   have  $\Delta = \frac{3}{2}\,r$. This implies $r+1=\frac{2}{3}$ and  once again 
$r = -\frac{1}{3}$. 
The chiral anomaly coefficients 
$\text{Tr}(\RR)$ and $\text{Tr}(\RR^3)$ receive contributions from all the fields  except  the vectors 
and thus we find from  \rf{2.11},\rf{3.4}
\begin{align}
\la{3.20}
\text{Tr}(\RR) &=\underbrace{-20\,r}_{\Psi_\mu}
\underbrace{+8\,(r+1)}_{T^+}
\underbrace{+r}_\psi = -20\, (-\tfrac{1}{3})+8\,(\tfrac{2}{3})-\tfrac{1}{3} = \tfrac{35}{3}, \notag \\
\text{Tr}(\RR^{3}) &=\underbrace{4\,r^{3}}_{\Psi_\mu}
\underbrace{-4\,(r+1)^{3}}_{T^+}
\underbrace{+r^{3}}_\psi = 4\, (-\tfrac{1}{3})^{3}-4\,(\tfrac{2}{3})^{3}+(-\tfrac{1}{3})^{3} = 
-\tfrac{37}{27}.
\end{align}


\subsubsection{Conformal graviton multiplet $[2]$}

The  superfield  description   of the   linearized $\N=1$  conformal supergravity multiplet $[2]$
 was  discussed in \cite{Ferrara:1977mv,Bergshoeff:1980is}. The corresponding real superfield 
   starting with  graviton contains (in a  Wess-Zumino gauge) the  components 
   corresponding to the  fields in \rf{33.1} 
\be
\la{3.21}
\def\arraystretch{1.3}
\begin{array}{cccc}
\toprule
& \Delta & (j_1, j_2) & r  \\
\midrule
h_{(\alpha\beta)(\dot\alpha\dot\beta)} & 0 & (1,1) & r		\\
\Psi_{(\alpha\beta)\dot\beta} & \frac{1}{2} & (1,\frac{1}{2}) & r+1 \\
\overline\Psi_{(\dot\alpha\dot\beta)\beta} & \frac{1}{2} & (\frac{1}{2},1) & r-1 \\
V_{\alpha\dot\alpha} & 1 & (\frac{1}{2},\frac{1}{2}) & r \\
\bottomrule
\end{array}
\ee
The expected value of  the graviton R-charge  is $r=0$. To   confirm this  one may consider the 
corresponding   chiral  field   strength superfield  \cite{Siegel:1980bp,Howe:1981gz}  (cf. \rf{3.19})
\be
\la{4.11}
W_{\alpha\beta\gamma} = \Psi_{\alpha\beta\gamma} + \theta^\delta
( C_{\alpha\beta\gamma\delta} + \eps_{\delta(\alpha}  F_{\beta\gamma})
   + \theta^2    \Phi_{\alpha\beta\gamma} \ , 
\ee
where $\Psi_{\alpha\beta\gamma}$   is the gravitino field  strength  and 
$\Phi_{\alpha\beta\gamma}$  is the "second"   gravitino field strength (self-dual part of the 
strength of $\Phi_\mu  \sim \gamma^\nu  \del_{[\m} \Psi_{\nu]} $).
Here $\Psi_{\alpha\beta\gamma}$  should have   $\Delta_{_\Psi} =\frac{3}{2}r_{_\Psi} = \frac{3}{2}$ 
 so  that 
 $r_{_\Psi} =1$ and  thus in \rf{3.21}    we should   have  $r=0$ (equivalently, 
 this follows from the fact that Weyl tensor  $C_{\alpha\beta\gamma\delta} $  has $r_{_C}= r_{_\Psi}  -1=0$). 

The chiral anomaly coefficients   $\text{Tr}(\RR)$ and $\text{Tr}(\RR^3)$ 
here receive contributions only from the Weyl   gravitino (see  \rf{2.11},\rf{3.4})
\begin{align}
\la{3.22}
\text{Tr}(\RR) =\underbrace{-20\,(r+1)}_{\Psi_\mu}= -20,\qquad\qquad 
\text{Tr}(\RR^{3}) =\underbrace{4\,(r+1)^{3}}_{\Psi_\mu}= 4 \ . 
\end{align}
These  values,   like those in \rf{3.12} and \rf{3.20},   are once again 
in agreement with the corresponding  values in   Table \ref{tab:multiplets}
demonstrating consistency with the supersymmetry  which  underlies  the relations \rf{210}.

\def \g {\gamma} 
\section{The superconformal index of $\mc N=1$ multiplets}
\la{sec:indices}

The explicit evaluation of the index \rf{2.1} 
 in a free  superconformal theory can be done in terms of the 
plethystic exponential\footnote{Eq.  (\ref{4.1}) is a standard way to build symmetric multi-particle states in terms
of the single-particle states. This is made explicit by the illustrative relation
\be
\notag 
\exp\sum_{n=1}^\infty\sum_{m=1}^\infty p^{n\g_m} = 1 + \sum_{m=1}^\infty p^{\g_m}
+\sum_{m\le m'} p^{\g_m+\g_{m'}}+\sum_{m\le m'\le m''} p^{\g_m+\g_{m'}+\g_{m''}}+\dots\, .
\ee
}  
\be
\la{4.1}
\log\text{I}(p,q) = \sum_{n=1}^\infty\frac{1}{n}\,\text{i}(p^n, q^n),
\ee
where the single-particle index $\text{i}(p,q)$ can be computed by letter 
counting \cite{Romelsberger:2005eg,Kinney:2005ej,Romelsberger:2007ec}. 
We shall  consider the special case of $p=q=e^{-\beta}\equiv t$ corresponding to \rf{2.4} 
and use the notation, cf. (\ref{2.4}), 
\be \la{4.10} \text{I}(e^{-\b},e^{-\b})  \equiv \text{I}(\b), \qquad \qquad 
 \text{i}(e^{-\b},e^{-\b})  \equiv \text{i}(\b) \ . \ee
Below we    will first  compute  the single-particle index $\text{i}(\b)$
for the  multiplets   introduced in the previous section  and then discuss  the   $\b\to 0$ expansion. 

\def \g {\gamma}

\subsection{Computing the single-particle  index}

\def \ti {{\textrm{i}}}

For the familiar   unitary  multiplets $[0]$ and $[1]$ one finds  \cite{Dolan:2008qi} 
\begin{align}
\la{4.2}
\text{i}_{[0]}(\beta) = 
\frac{   t^{ \frac{2}{3} }      -   t^{ \frac{4}{3}  }}{(1-t)^2},\qquad \qquad 
\text{i}_{[1]}(\beta) = -\frac{2\,t}{1-t} \ , \qquad \qquad t\equiv e^{-\b} \ . 
\end{align}
For the $[0']$ multiplet in \rf{3.9}  the analysis goes as follows. One has to consider  the fields   $X$  ("letters")
with $\delta \equiv  \Delta-2\,j_2-\frac{3}{2}\,r=0$ contributing 
\be
\la{4.3}
\delta(X)=0:\qquad \left. \textrm{i}\right|_X = (-1)^F\,t^{2j_2+r} = (-1)^F\,t^{\Delta-\frac{r}{2}} \ .
\ee 
Applying derivatives $\partial_{\alpha\dot\alpha}$ one 
builds new letters. 
In the following it will be convenient to  denote 
 the spinor  indices $\alpha=1,2$ by $\pm$:   $1\to +$ and $2\to -$  (and similar for dotted indices).   The $+/-$ notation is convenient because 
each type of index (dotted or undotted) with such a value increases/decreases the third component of the associated Lorentz spin by $\frac{1}{2}$.
The  derivatives
$\partial_{\pm \dot +}$ do not change $\delta$. 
They can be applied repeatedly leading to a universal
factor $1/(1-t)^2$ in the  single-particle  index $\ti$.  
Instead, $\partial_{\pm \dot -}$ increase $\delta$ by 
two units. Any derivative $\partial_{\alpha\dot\alpha}$ does  not change $r$ so, on the $\delta=0$ states,  
it increases 
$2j_2+r \stackrel{\delta=0}{=} \Delta-\frac{r}{2}$ by
one unit. This gives the set  of contributions in  Table~\ref{tab:0}.
\begin{table}
\begin{center}
\begin{tabular}{ccccc}
\toprule
& $\Delta$ & $(j_{1}, j_{2})$ & $r$   &  
$(-1)^F\,t^{2j_2+r}$\\
\midrule
$\pf $ & 0 & (0,0) & 0  & 1 \\
\midrule
$\overline{\phi}^{(2)}$ & 0 & (0,0) & 0 &   1 \\
$\overline\psi^{(3)}_{\dot -}$ & $\frac{1}{2}$ & $(0,-\frac{1}{2})$ & 1 &   -1 \\
$\partial_{\pm \dot -}\overline\psi^{(3)}_{\dot +}$ & $\frac{3}{2}$ & $(\pm\frac{1}{2},0)$ & 1 &   
$-2t$ \\
$\partial_{\pm \dot -}\overline{\phi}$ & 2 & $(\pm\frac{1}{2},-\frac{1}{2})$ & 2 &   $2t$\\
\bottomrule
\end{tabular}
\end{center}
\caption{Contributions to the index of the $[0']$ multiplet.  $j_{1}$ and $j_{2}$ (which can take positive and negative  values) stand for 
third components of the two $SU(2)$   spins 
 which  label the states  in   the  superconformal  index. 
 Each undotted (dotted) 
$\pm$ index contributes $\pm\frac{1}{2}$ to $j_{1}$($j_{2}$).}
\la{tab:0} 
\end{table}
The descendants (obtained by the   application of $\partial_{\pm \dot +}$  
leaving $\delta$ invariant)
that have the form of 
equations of motion are  
\begin{align}
& \partial_{- \dot +} (\partial_{+ \dot -}\overline\phi) \sim\Box\overline\phi=0\ , \la{4.4} \\
& \partial_{\pm \dot +}\partial_{- \dot +} (\partial_{+ \dot -}\overline\psi^{(3)}_{\dot +})
\sim\Box\partial_{\pm \dot +}\overline\psi_{\dot +}^{(3)}=0\ .  \la{4.5}
\end{align}
This gives the index  (cf. \rf{4.2})  
\be
\la{4.6}
\text{i}_{[0']}(\b) = \frac{1+1-1-2\,t+(2\,t-t^2)}{(1-t)^2} = 
 \frac{1-t^2}{(1-t)^2} \ .
\ee
A similar analysis can be carried out  for the tensor multiplet $[{1\ov 2}]$ in \rf{3.11}. In this case,
 the list of non-zero contributions
is collected in Table \ref{tab:tensor}.
\begin{table}
\begin{center}
\begin{tabular}{ccccc}
\toprule
& $\Delta$ & $(j_{1}, j_{2})$ & $r$  & $(-1)^F\,t^{2j_2+r}$\\
\midrule
$\psi^{(3)}_\pm$ & $\frac{1}{2}$ & $(\pm\frac{1}{2},0)$ & $\frac{1}{3}$ &  $-2\,t^\frac{1}{3}$ \\
\midrule
$\overline\psi^{(3)}_{\dot +}$ 
& $\frac{1}{2}$ & $(0,+\frac{1}{2})$ & $-\frac{1}{3}$ &  $-t^\frac{2}{3}$\\
$\overline{\phi}$ & 1 & $(0,0)$ & $\frac{2}{3}$ & $t^\frac{2}{3}$ \\
$T_{\dot +\dot -}^-$ & 1 & $(0,0)$ & $\frac{2}{3}$ &  $t^\frac{2}{3}$ \\
$\partial_{\pm \dot -}  T^-_{\dot +\dot +}$ & 2 & $(\pm\frac{1}{2},\frac{1}{2})$ &$ \frac{2}{3}$  & 
$2\,t^\frac{5}{3}$ \\
$\overline\psi_{\dot -}$ & $\frac{3}{2}$ & $(0,-\frac{1}{2})$ & $\frac{5}{3}$ &  $-t^\frac{2}{3}$ \\
$\partial_{\pm \dot -}\overline\psi_{\dot +}$ & $\frac{5}{2}$ & $(\pm\frac{1}{2},0)$ & $\frac{5}{3}$ &  
$ -2\,t^\frac{5}{3}$\\
\bottomrule
\end{tabular}
\end{center}
\caption{Contributions to the index of the $[\frac{1}{2}]$ multiplet. 
The components of the 
(anti) self-dual tensor are symmetric in the (dotted) undotted indices.
}
\la{tab:tensor} 
\end{table}
The contribution of the  last line of this table should not be  included:
 due to the spinor equations of motion, this  derivative   may
be replaced by another one with $\delta\neq 0$. This gives the contributions 
$-2\,t^\frac{1}{3}$ and $+2\,t^\frac{5}{3}$ from the  "left" and "right" chiral fields.
Next, we have to take into account the equations of motion for the $(j_1,j_2)=(1,0)$
and $(j_1,j_2)=(0,1)$ chiral components. One can check that there are no contributions with $\delta=0$  and  thus 
\be
\la{4.7}
\text{i}_{[\frac{1}{2}]}(\b) = \frac{-2\,t^\frac{1}{3}+2\,t^\frac{5}{3}}{(1-t)^2}.
\ee
To find the index  for  the gravitino multiplet $[\frac{3}{2}]$ in \rf{3.18} 
we  should  take into account that  relevant letters should be gauge invariant, {\em i.e.}
  use  the gravitino field strength in  \rf{3.19}  and also the 
"second" gravitino field strength 
$\Phi_{\alpha\beta\gamma}$  (cf. \rf{4.11}).
The  latter    obeys  the Bianchi identity
\be
\la{4.8}
\partial\indices{_\alpha^{\dot \beta}}
\partial\indices{_\beta^{\dot \gamma}}
\overline\Psi_{\dot\alpha\dot\beta\dot\gamma} = \partial\indices{_{\dot\alpha}^\gamma}
\Phi_{\alpha\beta\gamma},
\ee
and thus  its  dimension  is $\frac{5}{2}$ while the R-charge is opposite to that of $\Psi_{\alpha
\beta\gamma}$. The  non-zero contributions are collected in Table \ref{tab:gravitino}.
\begin{table}
\begin{center}
\begin{tabular}{ccccc}
\toprule
 & $\Delta$ & $(j_{1}, j_{2})$ & $r$  &  $(-1)^F\,t^{2j_2+r}$\\
\midrule
$T_{\alpha\beta}^+$ & 1 & $(j_{1},0)_{j_{1}= 0, \pm 1}$ & $\frac{2}{3}$ &  $3\,t^{\frac{2}{3}}$ \\
$\FF_{\alpha\beta}$ & 2 & $(j_{1},0)_{j_{1}= 0, \pm 1}$ & $\frac{4}{3}$ & $3\,t^\frac{4}{3}$\\
\midrule
$T^-_{\dot +\dot +}$ & 1 & $(0,+1)$ & $-\frac{2}{3}$ &  $t^{\frac{4}{3}}$  \\
$\overline\Psi_{\dot +\dot +\dot -}$ &$\frac{3}{2}$ & $(0,+\frac{1}{2})$ & 
$\frac{1}{3}$
&  $-t^{\frac{4}{3}}$  \\
$\partial_{\pm \dot -}\overline\Psi_{\dot +\dot +\dot +}$ & $\frac{5}{2}$ &
$(\pm\frac{1}{2},1)$ & 
$\frac{1}{3}$ &   $-2\,t^{\frac{7}{3}}$ \\
$\overline\Phi_{\dot +\dot +\dot +}$ & $\frac{5}{2}$ & $(0,+\frac{3}{2})$ & 
$-\frac{1}{3}$ &  $-t^\frac{8}{3}$ \\
$\overline\psi_{\dot +}$ & $\frac{3}{2}$ & $(0,+\frac{1}{2})$ & $\frac{1}{3}$ &  
$-t^{\frac{4}{3}}$ \\
$\overline{\FF}_{\dot + \dot -}$ & 2 & $(0,0)$ & $\frac{4}{3}$ &  $t^{\frac{4}{3}}$ \\
$\partial_{\pm \dot -}\overline {\FF}_{\dot +\dot +}$ & 3 & 
$(\pm\frac{1}{2},+\frac{1}{2})$ & $\frac{4}{3}$ &  
$2\,t^{\frac{7}{3}}$
 \\
\bottomrule
\end{tabular}
\end{center}
\caption{Contributions to the index of the $[\frac{3}{2}]$ multiplet.  
}
\la{tab:gravitino} 
\end{table}
The  Bianchi identity  for  the (complex) Maxwell field strength
\be
\la{4.9}
\partial\indices{_\alpha^{\dot \beta}}\,\overline \FF_{\dot \alpha\dot \beta} = 
\partial\indices{^\beta_{\dot \alpha}}\,\FF_{\alpha\beta}
\ee
lead  to additional   vector contribution   $-2\,t^{\frac{4}{3}+1}-2\,t^\frac{7}{3}= -4\,t^\frac{7}{3}  $. 
Finally, we   should  account for  the
equations of motion of $T^+_{\alpha\beta}$, {\em i.e.} 
$\partial_{\alpha\dot \alpha}\partial_{\beta\dot \beta}\,T^{+\alpha\beta}=0$  getting  another 
$-t^{\frac{2}{3}+2}$ contribution.  As a result, the  final expression for the index  is 
\be
\la{4.100}
\text{i}_{[\frac{3}{2}]}(\beta) = \frac{-2 t^\frac{8}{3}-4\, t^\frac{7}{3}+3\, t^\frac{4}{3}+3\, 
t^\frac{2}{3}}{(1-t)^2}.
\ee
For the graviton multiplet $[2]$ in \rf{3.21} the 
computation  of the index should  be again  done in terms of the gauge invariant field strengths  appearing  in  \rf{4.11}. 
The resulting non-zero contributions  are collected in Table \ref{tab:graviton}.
\begin{table}
\begin{center}
\begin{tabular}{ccccc}
\toprule
& $\Delta$ & $(j_{1}, j_{2})$ & $r$  &  $(-1)^F\,t^{2j_2+r}$\\
\midrule
$\Psi_{\alpha\beta\gamma}$ & $\frac{3}{2}$ & $(j_{1},0)_{j_{1}=\pm\frac{1}{2},\pm\frac{3}{2}}$ 
& 1 & $-4\,t$ \\
\midrule
$\overline\Psi_{\dot +\dot +\dot +}$ & $\frac{3}{2}$ & $(0,+\frac{3}{2})$ & -1 
&  $-t^2$ \\
$\overline\Phi_{\dot +\dot +\dot -}$ & $\frac{5}{2}$ & $(0,+\frac{1}{2})$ & 1 &
$-t^2$ \\
$\partial_{\pm \dot -}\overline\Phi_{\dot +\dot + \dot +}$ & $\frac{7}{2}$ &
$(\pm\frac{1}{2},+1)$ & 1 &
 $-2\,t^3$ \\
$\overline C_{\dot +\dot +\dot +\dot -}$ & 2 & $(0,+1)$ & 0 &  $t^2$  \\
$\partial_{\pm \dot -}\overline C_{\dot +\dot +\dot +\dot +}$ & 3 & 
$(\pm\frac{1}{2},+\frac{3}{2})$ & 0 &  $2\,t^3$ \\
$\overline F_{\dot+\dot +}$ & 2 & $(0,+1)$ & 0 &  $t^2$ \\
\bottomrule
\end{tabular}
\end{center}
\caption{Contribution to the index of the $[2]$ multiplet.}
\la{tab:graviton} 
\end{table}
Before taking into account Bianchi identities, the index is simply $\frac{-4 t}{(1-t)^2}$.
The Bianchi identities may only contribute
a term proportional to $t^3$. The condition of vanishing of the index numerator for $t=1$  then gives\footnote{
This extra $+4\,t^3$ comes from the  field in the 4th line of the Table \ref{tab:graviton} 
 (which enters a Bianchi identity
contributing $+2t^3$) and from  $\partial\indices{^\a_{\dot \a}}\partial\indices{^\b_{\dot \b}}
\Psi_{\a\b\g} + \dots=0$ (which is another Bianchi identity similar  to \rf{4.8} conserving  $\delta=0$ when 
$\dot \a = \dot \b = -$ and $\g=\pm$ is arbitrary).}
\be\la{4.12}
\text{i}_{[2]}(t) = \frac{-4 t+4\,t^3}{(1-t)^2}.
\ee
The  summary  of the computed indices is presented  in Table \ref{tab:indices}.\footnote{
The supersymmetric index is the $n=1$ case of similar indices for  
theories on the lens space $S^1_\beta\times S^3/\mathbb{Z}_n$ that have been computed
in \cite{Benini:2011nc} for unitary theories. In the $n >1$  cases the index receives 
contributions from twisted sectors. 
It would be interesting to extend the analysis to the non-unitary multiplets considered here
and explore the $n\to \infty$ limit 
when $S^3/\mathbb{Z}_n\to S^2$ and the index reduces to that of a 3d theory.}
\begin{table}
\begin{center}
\begin{tabular}{ccccccc}
\toprule
& $[0]$ & $[1]$ & $[0']$ & $[\frac{1}{2}]$ & $[\frac{3}{2}]$ & $[2]$ \\
\midrule
$P(t)$ & $t^\frac{2}{3}-t^\frac{4}{3}$ & $-2t+2t^2$ & $1-t^2$ & $-2\,t^\frac{1}{3}+2\,t^\frac{5}{3}$ & 
$-2 t^\frac{8}{3}-4\, t^\frac{7}{3}+3\, t^\frac{4}{3}+3\, 
t^\frac{2}{3}$ & $-4t+4t^3$  \\
\bottomrule
\end{tabular}
\end{center}
\caption{Numerators $P(t)$   of  the 
single particle indices  $\text{i}(\beta) = \frac{P(t)}{(1-t)^{2}}$ 
of the $\mc N=1$ multiplets. 
}
\la{tab:indices} 
\end{table}

\def \RI  {{\rm I}}\def \ri  {{\rm i}}
\def \ZZ {{\rm  z}} 

\subsection{Small $\beta$ expansion of  the index ${\rm I}(\b)$} 
\la{sec:zeta}

Let us   now  use the above results  for $\ri(\b)$   to compute   the small $\beta$ expansion of the superconformal
index ${\rm I}(\b)$  in order to compare with the expected   expansion (\ref{2.6}).
A generalization to a 2-parameter family of 
unequal $p$ and $q$ in \rf{2.1} will be discussed in Appendix \ref{app:squashed}.

 The usual approach to  derivation of the small $\beta$ expansion  of the index 
for models involving unitary multiplets starts from the 
summation in (\ref{4.1}) in terms of elliptic $\Gamma$ function. Modular properties 
of the resulting expressions \cite{felder2000elliptic} are then exploited to discuss the small $\beta$ limit. 
Here we propose  a simpler approach based on the 
 techniques developed for studying similar  
limit of  standard partition functions  using that 
  ${\rm I}(\b)$   has a formal structure of a 
 partition function. Let  $\text{m}$   be a  label a  particular multiplet  and   let us 
 define 
 the Mellin transform of the single-particle  index $\ri(\b)$ as 
\be
\la{4.13}
\ZZ_\text{m}(u) \equiv  \frac{1}{\Gamma(u)}\int_0^\infty\,d\beta \, \beta^{u-1}\,\text{i}_\text{m}(\beta) \ . 
\ee
Then for $\RI(\b)$ in \rf{4.10} one finds\footnote{Eq.~(4.14) follows from the Mellin inversion formula and may be checked 
for a typical single-particle contribution to the single-particle index starting
from 
 the  relation $e^{-\beta}=\frac{1}{2\pi i}\int_{v-i\infty}^{v+i\infty} du\, 
\beta^{-u}\,\Gamma(u)$ valid for $v>0$ (see, {\em e.g.}, \ci{Gibbons:2006ij}).  Here $\zeta(u+1)$ is the Riemann zeta function. 
Note that the  relations  below apply to both fermions and bosons (as fermions are treated as periodic on the circle). }
\be
\la{4.14}
\log \text{I}_\text{m}(\beta) =    \sum_{n=1}^\infty\frac{1}{n}\,\text{i}_\text{m}(n \b)    
=     \frac{1}{2\pi\,i}\int_{v-i\infty}^{v+i\infty}
\,du\ \beta^{-u}\,\Gamma(u)\,\zeta(u+1)\,\ZZ_\text{m}(u)\ ,
\ee
which is valid when $v$ is sufficiently large.
When the vertical contour in (\ref{4.14}) is moved to the left, we pick up residues
of poles at integer $u$  and this has the form of a  small $\beta$ 
expansion with the  coefficients 
involving  the residues of $\ZZ_\text{m}(u)$.\footnote{In some cases
symmetry properties of the integrand allow one  to relate the contour associated with $-v$ to that 
at $+v$. Then  the remainder of the small $\b$ (high temperature) pole expansion can be found explicitly 
and gives rise to a "temperature inversion" relation \cite{Gibbons:2006ij}.}
In general, 
given a term $\frac{t^q}{(1-t)^2}$ in the index, we may   use  that 
\begin{align} 
\la{4.20}
 \sum_{n=0}^\infty \frac{1}{\Gamma(u)}(n+1)\int_0^\infty d\beta\,\beta^{u-1}\,e^{-n\,\beta}
e^{-q\,\beta} &= \sum_{n=0}^\infty(n+1)(n+q)^{-u}\notag \\
&= \zeta(u-1,  q)+(1-q)\,\zeta(u, q),
\end{align}
and taking residues in (\ref{4.14}) we immediately obtain the expansion of the index. 


Let us apply this method to the known cases of  the chiral and vector multiplets.
 For the $[0]$ chiral multiplet we have from  (\ref{4.2}) 
\begin{align}
\la{4.15}
\ZZ_{[0]}(u) &= \frac{1}{\Gamma(u)}\int_0^\infty d\beta\,\beta^{u-1}\frac{e^{-{2\ov 3} \beta}  -  e^{-{4\ov 3} \beta}   }{(1-e^{-\beta})^2} 
= \frac{1}{\Gamma(u)}\sum_{n=0}^\infty(n+1)
\int_0^\infty d\beta\,\beta^{u-1}e^{-n\,\beta}\ (e^{-{2\ov 3} \beta}  -  e^{-{4\ov 3} \beta} ) \notag \\
&= \sum_{n=0}^\infty(n+1)\,\Big[(n+\tfrac{2}{3})^{-u}-(n+\tfrac{4}{3})^{-u}\Big]
\no \\ &
= \zeta (u-1,\tfrac{2}{3})-\zeta(u-1,\tfrac{4}{3})+
\tfrac{1}{3}\,\zeta(u,\tfrac{2}{3})+\tfrac{1}{3}\,\zeta(u,\tfrac{4}{3}).
\end{align}
Multiplying this by  $\Gamma(u)\zeta(u+1)\beta^{-u}$ (to get the integrand in  (\ref{4.14}))
 and taking the residues
of the poles, we get  contributions to (\ref{4.14})   coming    from $u=-1,0,1$ only. As a result,  
\be
\la{4.16}
\log \text{I}_{[0]}(\beta) = \frac{\pi^2}{9\,\beta}+\text{k}[0]+\frac{7}{324}\,\beta + \mc O(e^{-1/\beta}),
\ee
where
\be
\la{4.17}
\text{k}[0] = \frac{\pi }{9 \sqrt{3}}-\frac{1}{6} \log  3 -\frac{\psi ^{(1)}\left(\frac{1}{3}\right)}{6
   \sqrt{3} \pi }\ . 
\ee
This is in full agreement with (\ref{2.6}) with the proposed values of the coefficients $C_i$,
see (\ref{2.3}), (\ref{2.7}),\rf{2.111},(\ref{2.12})  and Table \ref{tab:multiplets}.  
 The constant $\text{k}[0]$ 
can be identified with the 
3d partition function $\log Z^{\rm susy}_{S^3}$. \footnote{\la{foot:ell} 
According to  \cite{Jafferis:2010un}  
$\log Z^{\rm susy}_{S^3} = \ell({1\ov 3})$,
where
$
\ell(R) = -R\,\log(1-e^{2\,\pi\,i\,R})-\frac{1}{2\pi\,i}\text{Li}_2(e^{2\,\pi\,i\,R})
+\frac{i\,\pi\,R^2}{2}-\frac{i\,\pi}{12}$.
It is possible to prove that  $\text{k}[0] = \ell({1\ov 3})$. The relation  
of the index to  3d partition function in the $\beta\to 0$ limit
 after the removal of singular terms is  a non-trivial fact depending on regularization, 
  see \cite{Dolan:2011rp,Gadde:2011ia,Imamura:2011uw,Niarchos:2012ah,
Agarwal:2012hs,Aharony:2013dha}. For a discussion of  this relation  in the  case of 
non-supersymmetric conformal
partition functions see Appendix \ref{app3d}.
}
Similarly, for a vector multiplet $[1]$ we find using  (\ref{4.2})
\begin{align}
\la{4.18}
\ZZ_{[1]}(u) &= -\frac{2}{\Gamma(u)}\int_0^\infty d\beta\,\beta^{u-1}\frac{e^{-\beta}}
{1-e^{-\beta}} = \notag \\
&= -\frac{2}{\Gamma(u)}\sum_{n=0}^\infty
\int_0^\infty d\beta\,\beta^{u-1}e^{-(n+1)\,\beta} =
-2\, \sum_{n=0}^\infty(n+1)^{-u} = -2\,\zeta (u)\ .
\end{align}
Taking residues in (\ref{4.14}), we obtain 
\be
\la{4.19}
\log \text{I}_{[1]}(\beta) = -\frac{\pi^2}{3\,\beta}-\log\beta + \log (2 \pi) 
+\frac{1}{12}\,\beta+ \mc O(e^{-1/\beta}) \ . 
\ee
This is again in agreement with (\ref{2.6})  and  \rf{2.3},(\ref{2.7}),(\ref{2.12}) and  Table \ref{tab:multiplets}. Eq.
(\ref{4.19}) implies that    $\text{k}[1] =  \log Z^{\rm susy}_{S^3}=\log(2\pi)$.
Consistency of this result is further discussed in App.~(\ref{app3d-susy}).

\

Applying the same  method  also  to the four non-unitary multiplets  with single-particle indices 
given in Table \ref{tab:indices}  the final  results  can be summarized  as follows 
\footnote{\la{foot:phi2} Let us note  that for the $[0']$ multiplet one has to be careful with the contribution of the  
higher derivative scalar $\pf $. This field has canonical dimension 0
(like a scalar in 2d) 
 and one finds terms
of the form $\sum_{n=0}^\infty n^{-u}$. The  $n=0$  term is ambiguous and 
we used the natural analytical continuation $0^{-u}\equiv 0$ for 
all (complex) $u$. 
}
\begin{align}
\la{4.21}
[0] &= (2\,\phi, \psi):\quad  
& \log \text{I}_{ [0] } (\beta) &= \frac{\pi^2}{9\,\beta}+\text{k}[0]+
0\cdot\log\beta+\frac{7}{324}\,\beta + \mc O(e^{-1/\beta}), \notag \\
[1] &= (V_\mu, \psi):\quad 
& \log \text{I}_{ [1] } (\beta) &= -\frac{\pi^2}{3\,\beta}+\text{k}[1] -  \log\beta
+\frac{1}{12}\,\beta+ \mc O(e^{-1/\beta}), \notag \\
[0'] &= (2\,\pf , \psi^{(3)}, 2\,\phi): \quad  
& \log \text{I}_{ [0'] }(\beta) &= \frac{\pi^2}{3\,\beta}+\text{k}[0']
+0\cdot \log\beta
-\frac{1}{12}\,\beta+ \mc O(e^{-1/\beta}), \notag \\
[\tfrac{1}{2}] &= (\psi, 2\varphi, T_{\mu\nu}, \psi^{(3)}):\quad
& \log \text{I}_{ [\frac{1}{2}] }(\beta) &= -\frac{4\,\pi^2}{9\,\beta}
+\text{k}[\tfrac{1}{2}]+0\cdot\,\log\beta
-\frac{1}{81}\,\beta
+ \mc O(e^{-1/\beta}), \notag \\
[\tfrac{3}{2}] &= (\Psi_\mu, 2\,V_\mu, T_{\mu\nu}, \psi):\quad
& \log \text{I}_{[\frac{3}{2}]}(\beta) &= \frac{13\,\pi^2}{9\,\beta}
+\text{k}[\tfrac{3}{2}]+6\,\log\beta
-\frac{389}{324}\,\beta
+ \mc O(\beta^2), \notag \\
[2] &= (h_{\mu\nu}, \Psi_\mu, V_\mu):\quad 
& \log \text{I}_{[2]}(\beta) &= -\frac{4\,\pi^2}{3\,\beta}+\text{k}[2]- 8\,\log\beta
+\frac{7}{3}\,\beta
+\mc O(\beta^2)\ , 
\end{align}
where 
\begin{align}
\no
&\text{k}[0]  = \frac{\pi }{9 \sqrt{3}}-\frac{1}{6} \log
   3 -\frac{\psi ^{(1)}\left(\frac{1}{3}\right)}{6
   \sqrt{3} \pi }\  , \qquad \qquad 
   \text{k}[1] = \log (2 \pi) \ ,\no \\
    &     \text{k}[0']=   0\  ,  \qquad \qquad 
\text{k}[\tfrac{1}{2}] = \frac{2\pi}{9\sqrt{3}}+\frac{2}{3}\log 3 
-\frac{\psi^{(1)}(\frac{1}{3})}
{3\sqrt{3}\,\pi}, \notag \\
&\text{k}[\tfrac{3}{2}] = \frac{\pi}{9\sqrt{3}}-\frac{49}{6}\log 3 
-\frac{\psi^{(1)}(\frac{1}{3})}
{6\sqrt{3}\,\pi}, \qquad \qquad 
\text{k}[2] = 4\,\log(2\pi).   \la{4.22}
\end{align}
These constant terms will be further discussed in Appendix \ref{app3d-susy}.\foot{Let us note   that  
part of the expansions in (\ref{4.21}) can be  found by a  naive  procedure of  
first expanding  the single-particle index 
and then applying (\ref{4.1}) term by term. In general, $\text{i}(\beta) = 
\frac{A_{-1}}{\beta}+A_0+A_1\,\beta+A_2\,\beta^2+\dots$. From (\ref{4.1}), we then  formally obtain:
$\log\text{I}(\beta) = \zeta(2)\,\frac{A_{-1}}{\beta}+\zeta(1)\,A_0+\zeta(0)\,A_1\,\beta+\zeta(-1)\,
A_2\,\beta^2+\dots = \frac{A_{-1}}{6}\,\frac{\pi^2}{\beta}+\zeta(1)\,A_0-\frac{1}{2}\,A_1\,\beta
-\frac{1}{12}\,\beta^2+\dots$. 
 One can check that $\frac{A_{-1}}{6}$ 
is indeed the coefficient of the leading term in (\ref{4.21}) in all cases. The same agreement is found for the 
linear in $\beta$ term. The term proportional to $A_0$ is ill-defined  but a  heuristic 
replacement rule $\zeta(1)\to -\log\beta$   reproduces  indeed 
the $\log\beta$ term in (\ref{4.21}).  However, all  other  subleading corrections  are not 
captured   correctly by this procedure.
}

\def \Dn  {\nu}

\section{Structure of small $\beta$ expansion of the  index of non-unitary multiplets}
\la{sec:structure}


Let us now compare  the explicit  values of the coefficients 
   appearing  in the small $\b$  expansion 
\rf{2.5}  of the indices in \rf{4.21}    with their expected values  discussed in Section \ref{sec:review},
{\em i.e.} with
the previously suggested  relations  \rf{2.8},\rf{2.111},\rf{2.3}.\footnote{Here we will not  attempt 
 to  compare the expressions  in (\ref{4.22})   with  their expected \rf{2.111}   values  $\log  Z^{\rm susy}_{S^3} $ 
 since to compute  the latter requires  first  the construction  of the 
  explicit supersymmetric  Lagrangians for the non-unitary  multiplets  on $S^1\times S^3$ that 
  should contain extra  couplings to the R-symmetry gauge field background.
  Nevertheless, we remark that $\text{k}[\frac{1}{2}] = \frac{2}{3}\,\ell({2\ov 3})+\frac{8}{3}\ell({1\ov 3})$,
  where $\ell(R)$ was  defined in footnote \ref{foot:ell}.}
  We shall  denote   the true values   of the  coefficients   as ${\widehat C}_i$   
  with $C_i$   being the expected values:
\be   \la{5.1a}
\log \text{I}(\beta) \stackrel{\beta\to 0}{=} {\widehat C}_{1}   \,\frac{\pi^2}{\beta}+ {\widehat C}_{2} +
{\widehat C}_{3} \log {\b} + \dots \ . \ee

\def \CSG {{\rm CSG}}
\def \SYM {{\rm SYM}}

\medskip
\noindent {\bf Leading term} $\sim 1/\beta$:

\noindent
Comparing the values of the  coefficient of the $\pi^2/\b$ term  in  (\ref{4.21}) 
with their expected \rf{2.8}  values $C_1=   -\frac{1}{3}\, \text{Tr}(\RR)$   in  Table \ref{tab:multiplets}, we find  agreement for the $[0],\ [1], \ [0']$ multiplets   but discrepancies for the non-unitary multiplets 
$[\frac{1}{2}]$, $[\frac{3}{2}]$, $[2]$  containing the antisymmetric tensor or conformal gravitino:  
\begin{align}
&
{\widehat C}_{1} =C_1 - {1\ov 3} \nu=  -\frac{1}{3}\,\big[\text{Tr}(\RR)+\Dn\big] \ , \la{5.1}
\\
&
\la{5.2} \Dn_{[0]} = \Dn_{[1]} = \Dn_{[0']}  =0 \ , \qquad \qquad 
\Dn_{[\frac{1}{2}]} = 8, \quad
\Dn_{[\frac{3}{2}]} = -16,\quad
\Dn_{[2]} = 24.
\end{align}
Remarkably, the correction terms $\Dn_\text{m}$   are all  integer multiples of 8. 
For  the collection of multiplets appearing (\ref{3.1})
in the $\N$-extended conformal supergravities  we  then get 
$
\Dn_{_{\mc N=1}} = 24,\
\Dn_{_{\mc N=2}} = 8,\
\Dn_{_{\mc N=3}} = 
\Dn_{_{\mc N=4} }= 0 \ . 
$
The $\N=3$ and $\N=4$   conformal supergravities also have $\aa=\cc$  \ci{Fradkin:1985am}  or $\tr \RR=0$  and thus ${\widehat C}_1 = C_1=0 $.
    The same result is found also for $\mc  N=4$    vector multiplet or $[1] + 3[0]$, {\it i.e.}
\be
\la{5.3}
{\widehat C}_{1_{\mc N=1\,\CSG }} =- {4\ov 3},\quad
{\widehat C}_{1_{\mc N=2\,\CSG}} = - {2\ov 9},\quad
{\widehat C}_{1_{\mc N=3\,\CSG}} =
{\widehat C}_{1_{\mc N=4\,\CSG}} = 0, \quad 
{\widehat C}_{1_{\mc N=4\,\SYM}} = 0.
\ee
The reason why  the relation \rf{2.8}   between the $1/\b$   term  
and $\text{Tr}(\RR)$   suggested in \cite{DiPietro:2014bca}
 fails to be universal  may   be due to  the fact  that 
 the argument in  \cite{DiPietro:2014bca}   may not directly apply 
 to theories   containing  more complicated chiral fields (self-dual tensors, conformal  gravitions, etc.)
rather  than  just  the standard  Weyl fermions.\foot{Indications that there are 
subtleties  in  reconstruction of 3d effective actions 
 in  the case of   (non-conformal) gravitinos  appeared in  \cite{Loganayagam:2012zg,Chowdhury:2015pba}.}
One possibility is that in reconstructing 3d effective action by matching   anomalies  
there is an  integer-shift ambiguity 
in the coefficient of the 3d Chern-Simons term   used in  \cite{DiPietro:2014bca}  
leading in general  to the presence of the 
correction term $\nu$ in \rf{5.1}. 
Note that  a  shift   of  $\widehat C_1 $ from    its    value  $C_1$ in \rf{2.7} 
 was    also   discussed  for    non-abelian gauge theories   in   \cite{Ardehali:2015bla}.

Let us   note also that the  presence of  the correction $\nu$ 
 is    similar to what happens in  non-unitary  2d   CFT.
 In a generic CFT  the partition function $Z(\beta)$ is related to the density of states $\rho(E)\,dE$.
Writing the energies in terms of the conformal dimensions and the 2d central charge, 
$E=2\pi\,(\Delta+\overline\Delta-\frac{1}{12}\text{c})$, the modular invariance $Z(\beta)=Z(\beta^{-1})$
implies that the limit $\beta\to 0$ is related to the $\beta\to \infty$ one  in which 
 $Z$ is dominated by the lowest-energy states $Z(\beta\to\infty)\sim A\,\beta^{-\l}
e^{-\beta\,E_\text{min}}$. Here $\l$ is non-zero for gapless systems so that 
\cite{Itzykson:1986pk,Kutasov:1990sv}
\be
\la{ks.5.4}
Z(\beta) = \int dE\,\rho(E)\,e^{-\beta\,E} \stackrel{\beta\to 0}{\sim} 
A\,\beta^\l\, e^{\frac{\pi}{6\,\beta}\,\text{c}_\text{eff}}, \qquad 
\text{c}_\text{eff} = -\frac{6}{\pi}E_\text{min} = \cc-24\,\Delta_\text{min} ,
\ee
where
$\Delta_\text{min} \equiv  \frac{1}{2}\min(\Delta+\overline\Delta).
$
In unitary theories one has  $\Delta_\text{min}=0$ and $\text{c}_\text{eff} = \text{c}$.
Instead, in non-unitary theories, $\Delta_\text{min}$ is typically negative and $\text{c}_\text{eff}>\text{c}$.
This inequality may be violated in the case of   supersymmetric partition function
where fermions are  taken to be periodic and contribute
with a negative sign.   In non-unitary case 
 one has generically  \cite{Kutasov:1990sv} 
$
\text{c}_\text{eff}(\text{B})-\text{c}_\text{eff}(\text{F})\neq 
\text{c}(\text{B})-\text{c}(\text{F})
$.
The correction $\Delta_\text{min} $  in   $\cc_\text{eff}=\cc-24\, \Delta_\text{min}  
$  is analogous  to the parameter $\nu$  in \rf{5.1}. 
There are, however,   important differences: modular invariance is not available in general
and the role of the 2d central charge is played by $\text{c}-\text{a}$ (cf. \rf{2.7}).
 Nevertheless,  it is tempting to
relate the  presence of the  non-zero $\nu$ in \rf{5.1}  with 
 the existence of negative norm states in the case of non-unitary multiplets. 

\iffa 
 \footnote{
Can $\Dn_\text{m}$ be explained by the global anomaly mechanism treated (in non - conformal case) in 
\cite{Chowdhury:2016cmh} to explain the gravitino peculiarities observed in 
\cite{Chowdhury:2015pba}  ? Apparently no because what these papers do is to compute 
the coefficient of the dimensionally reduced 3d effective theory by matching global anomalies,
$Z\to \exp(i\pi\eta) Z$ under large diffeomorphisms, to anomalous term $\sim \int_{\mc 
M_3} A\wedge dA$. They can do this up to an intrinsic ambiguity of $\eta \equiv \eta+2$, but this
does not means that the effective action is ambiguous. Problem is how to compute it correctly
for $T$ and $\psi_\mu$.
}
\fi 

\def \Dg {\gamma}

\medskip
\noindent {\bf Logarithmic term} $\sim\log\beta$:

\noindent
The  comparison between the coefficients of $\log\beta$ in (\ref{4.21}) and the suggested 
values $C_3= -4\,(2\text{a}-\text{c})$  in (\ref{2.12}) \ci{Ardehali:2015hya}    given in Table \ref{tab:multiplets} 
again implies   the presence of an integer  correction: 
\begin{align} 
& {\widehat C}_3 = C_3   + \Dg = -4\,(2\text{a}-\text{c}) + \Dg \ ,\la{58b} \\
&
 \Dg_{[0]} = \Dg_{[1]} = 0\ ,  \qquad \qquad\qquad
 \Dg_{[0']}  =-1, \quad 
\Dg_{[\frac{1}{2}]} = -3, \quad
\Dg_{[\frac{3}{2}]} = 3 ,\quad
\Dg_{[2]} = -1 \ , \la{5.8}\\
&{\widehat C}_{3\, [0]} = {\widehat C}_{3\, [0']} = {\widehat C}_{3\, [\frac{1}{2}]} = 0 \ , 
\  \qquad  {\widehat C}_{3\, [1]} = -1 \ , \ 
 \ \ {\widehat C}_{3\, [\frac{3}{2}]} = 6 \ , \ \ {\widehat C}_{3\, [2]} = -8 \ . \la{5.88}
\end{align}
Considering  the collections   of multiplets appearing in 
$\N$-extended conformal supergravities (\ref{3.1})  one finds
\be
\la{5.88a}
{\widehat C}_{3_{\mc N=1\,\CSG }} =24,\quad
{\widehat C}_{3_{\mc N=2\,\CSG}} = 8,\quad
{\widehat C}_{3_{\mc N=3\,\CSG}} = 0,\quad
{\widehat C}_{3_{\mc N=4\,\CSG}} = 2, \quad 
{\widehat C}_{3_{\mc N=4\,\SYM}} = -1.
\ee
Instead of trying to understand   why   ${\widehat C}_3$ is, in general,  different  from $ C_3=-4\,(2\text{a}-\text{c})$  proposed  in \cite{Ardehali:2015hya} 
    let us suggest an alternative  general  expression  for it.
 Let us start by noting  that the logarithmic term in the expansion of  the index  is the same as in the expansion of the supersymmetric partition function \rf{2.5}  and
  thus  may  have  a universal  origin. One may attempt to   interpret  the   singular $\log \b$ term  
appearing in the $\b\to 0$ limit  as    associated with   the KK  modes that  become 
 "massless"   in the limit of shrinking $S^1$. 
 In  practice, this relation is not  straightforward and  depends on 
 regularization, see also Appendix  \ref{app3d}.\foot{Let us note  also  that the 
$\log\beta$ term 
  in  the case of unitary non-abelian gauge theories  was discussed in 
  \cite{Ardehali:2015bla} where its coefficient
 was  related to the dimension of the  space of  flat directions (with no curvature coupling)  in the 
3d theory.} 

One may  then   expect that  the coefficient of the $\log \b$ term  should   be the   same as in the  
standard  partition function for   the 
 conformal  gauge  fields on $S^1_\beta\times S^3$  (with both bosons and fermions 
  taken to be  periodic on the  circle). 
Then the  $\log \b$  term  should receive contributions only  from the conformal   gauge  fields in each multiplet. 
 These  can be   found  from 
  the  conformal higher spin partition functions derived in \cite{Beccaria:2014jxa}
  and reviewed in Appendix \ref{app:CHS}   below. 
  
  
  The analysis in  Appendix \ref{app:CHS}
shows that the $\log\beta$ contribution comes from a specific  $SO(4,2)$ conformal character and is
determined by a particular integer equal up to sign to the number $n_{_\text{CKT}}$
of   conformal Killing tensors  for  the bosons  and 
 the number $n_{_\text{CKS}}$ of conformal Killing spinor-tensors
for  the fermions.\foot{Note that the UV finite partition  function on $S^1_\b \times S^3$   where $S^3$ has  radius $R$ (set to 1 in the above discussion)  contains the $\log \b$ term as  part of the dimensionless 
 $\log (\b/R)$ term. This suggests (see also Appendix \ref{app3d})
   that  may be  the  dependence on $R$   coming from some    
  zero modes  
 is    determined by  $n_{_\text{CKT}}$. 
 The  total  power of $R$    depends on a regularization of the  contribution of all other modes
  (for a related  discussion on $S^3$   see \ci{Giombi:2013yva}). 
 }



We  propose that  the coefficients ${\widehat C}_3$ 
of the  $\log \b$ term   in the  expansion of the supersymmetric partition function on $S^1_\beta\times S^3$
for a generic superconformal multiplet  should be given  
by the sum of  the   contributions from  the conformal higher spin  gauge fields in this multiplet,  {\em i.e.} 
\be \la{513}
{\widehat C}_3 \equiv  - {\rm n} \ , \qquad \qquad {\rm n}
=\sum_i  n_{_\text{CKT}} (i) -  \sum_i n_{_\text{CKS}}(i)\ .
\ee
In the   case of the  multiplets   discussed in this paper  the relevant  conformal gauge fields are the  standard vector $V_\m$ ($s=1$), 
the conformal graviton $h_{\m\n}$ ($s=2$)  and   the conformal gravitino $\Psi_\m$ ($\s=1$)  for which we find from \rf{511},\rf{5.13}:
\be
\la{5.14}
n_{_\text{CKT}}(V) = 1, \qquad \qquad  n_{_\text{CKT}}(h) = 15, \qquad \qquad 
n_{_\text{CKS}}(\Psi) = 8\ .
\ee
As a result,  from \rf{33.1}   we get 
\begin{align}
&{\widehat C}_{3\, [0]} = 0 \ , \ \qquad  {\widehat C}_{3\, [1]} = -1 \ , \ \qquad 
{\widehat C}_{3\, [0']} =  0 , \qquad 
\ \ {\widehat C}_{3\, [\frac{1}{2}]} = 0  \ , \la{515}  \\
&  {\widehat C}_{3\, [\frac{3}{2}]} =2\times (-1) + 8= 6 \ ,\qquad \quad\
 {\widehat C}_{3\, [2]} =-1 - 15  +8=-8  \ ,\la{588}
 \end{align}
exactly in    agreement with (\ref{4.21})  and \rf{5.88}.

\medskip
\noindent
{\bf Linear term} $\sim\beta$:

\noindent
For this term   there is full agreement between the expected   values of the supersymmetric  energy (\ref{2.3}) 
in Table \ref{tab:multiplets}   and the  coefficients  in (\ref{4.21}). 

\medskip
\noindent
{\bf Higher order corrections}:

\noindent
Higher order corrections in the small $\beta$ expansion of $\log \RI(\b)$  can be found by 
taking residues in (\ref{4.14}) at the points $u=-2, -3, \dots$. For the 
 unitary  multiplets one  can check that (\ref{4.15}) as well as   (\ref{4.18}) 
have no poles at these points. This means that the corrections to the expansions in  (\ref{4.21}) 
are exponentially suppressed as $\beta\to 0$ (see \rf{4.16},\rf{4.18}).
 The same conclusion can be drawn by repeating the 
analysis for the "non-gauge" multiplets $[0']$ and $[\frac{1}{2}]$. 

Instead, for the   non-unitary   multiplets $[{3\ov2}]$ and $[2]$  containing gauge fields  we 
find, in addition to the  leading terms given in  (\ref{4.21}),   an infinite series of power corrections
\begin{align}
\la{5.15}
\log \text{I}_{[\frac{3}{2}]}(\beta) =& 
... -\frac{389}{324}\,\beta+\frac{\beta ^2}{18}-\frac{\beta ^4}{6480}+\frac{11 \beta
   ^6}{11022480}
   -\frac{43 \beta ^8}{5290790400}+\frac{19 \beta
   ^{10}}{261894124800}+\dots, \notag \\
\log \text{I}_{[2]}(\beta) = & 
...+\frac{7}{3}\,\beta-\frac{\beta ^2}{6}+\frac{\beta ^4}{720}-\frac{\beta
   ^6}{45360}
   +\frac{\beta ^8}{2419200}-\frac{\beta
   ^{10}}{119750400}+\dots.
\end{align}
A similar pattern is found for the non-supersymmetric partition function of the conformal higher   spin 
fields. For instance, in the case of a spin $s$ bosonic conformal  field  the correction to (\ref{5.11}) can be  written as \be
\la{5.16}
\log Z_s = \text{terms in (\ref{5.11})} + R_s(\beta) \ ,
\ee
where  $R_s(\beta)$ is a  infinite  series  that can be  found   in a closed form 
\begin{align}\la{519}
R_s(\beta) = \frac{1}{3}\,\sum_{k=1}^{s-1}
k(k+1)\big[(2s+1)\,k-3s^2-2s-1\big]\,\log\frac{2\,\sinh[ \frac{(s-k)}{2}\b]}{(s-k)\,\beta}\ .
\end{align}
This  vanishes  for the Maxwell field ($s=1$),  while  for $s >1$   one finds
\begin{align}\la{514}
R_s(\beta) =\te  -\frac{(s-1) s^2 (s+1)^2 (s+2) (2 s+1)}{2160}\,\beta^2
+
\frac{(s-1) s^2 (s+1)^2 (s+2) (2 s+1) \left(9 s^2+9
   s-26\right)}{7257600}\beta^4+\dots.
\end{align}

\medskip

To summarize, the above  analysis of  non-unitary $\mc N=1$ multiplets shows that in general 
(cf. \rf{2.6})
\be
\la{5.19}
\log\text{I}(\beta) \stackrel{\beta\to 0}{=} -\big[16\,(\text{a}-\text{c})+
\Dn\big]\,\frac{\pi^2}{3\beta} -  {\rm n}\,\log\beta + {\rm k} 
+\frac{4}{27}(\text{a}+3\text{c})\,\beta+R(\beta) \ . 
\ee
Here 
 $\Dn$ is a   integer multiple of 8 which is non-zero  for  non-unitary 
 multiplets  with higher spin fields 
 and ${\rm n}= {\widehat C}_3$     given by \rf{513} is another  integer    which is   non-zero only for  multiplets with 
gauge fields.
${\rm k} \equiv C_2$  is a constant that should  be related \rf{2.111} to the partition function on $S^3$ 
and $R(\beta)$  contains power-like corrections 
for multiplets  with conformal higher spin gauge   fields but is 
$\mc O(\b^k e^{-1/\b})$  otherwise. 

It is of interest to consider  special  combinations of the   basic    $\N=1$ multiplets $[0],[1]$ and $[2], [{3\ov 2}],  [{1\ov 2}],[0']$ 
that   have  vanishing  leading $\b^{-1}$ and $\b$ coefficients  in the expansion of the index \rf{5.19}, i.e. 
 have vanishing total     a, c  and $\nu$ coefficients
\begin{align} 
&\aa_{\rm tot} =\cc_{\rm tot} =0 \ , \qquad  \nu_{\rm tot}=0: \no \\
& 
\te [2] + (k+3) [{3\ov 2}] + k'[1] + (2k+3)  [{1\ov 2}]+ (22k+18) [0] +(k' -9k-11) [0'] \ . \la{519} 
\end{align} 
Here $k$ and $k'$ are integers   and  we assumed   that there is  just one graviton multiplet.
As  $k>-1$ (for the number of $[0]$ not to be  negative)
  the simplest  solutions are  $k=0$ and $k'=11,12, ...$ for which 
 there are 4  conformal gravitini   as in $\mc N=4$  conformal supergravity. 
 Using \rf{588}  we then find that the coefficient  n of the $\log \b$ term in \rf{5.19} is 
 n=$10-k'$. 
 
  The case with $k=0, \, k'=12$, i.e. 
  $\te [2] + 3 [{3\ov 2}] + 12 [1] + 3  [{1\ov 2}]+ 18 [0] + [0'] $, 
    corresponds to 
  the familiar case of the  $\mc N=4$  conformal supergravity   coupled to   four copies of $\mc N=4$ SYM multiplets
($[1]_4 = [1]_1+3[0]_1$) which is a superconformal theory 
not only  at the quadratic  but  also the interacting level \ci{Fradkin:1983tg,Fradkin:1985am}.
 The  small $\b$  expansion of the superconformal  index  of this theory (which  does not depend on the 
  conformal supergravity and the SYM coupling constants)  is  given by
  \be \la{520}
  \log\text{I}(\beta) \stackrel{\beta\to 0}{=}  -2\,\log\beta +  {\rm k}   +  \frac{\beta^4}{1080}+\mc O(\beta^6) \ , 
\ee
  where  
    the  infinite series of  power corrections  
come  only from the $\mc N=4$  conformal supergravity  contribution (cf. \rf{5.15}).

The  minimal  solution for the superconformal theory \rf{519}  having 
 $k=0, \, k'=11$   has  a smaller field    content  (it corresponds to removing  the pair 
$[0']+[1]$ that has zero  anomalies, see Table 2):
$\te [2] + 3 [{3\ov 2}] + 11 [1] + 3  [{1\ov 2}]+ 18 [0] $.
It    has similar   expansion of the  index: 
$\log\text{I}(\beta){=}  -\,\log\beta +  {\rm k}   +...$.  
This   combination  cannot be written  as a collection of $\N=3$ or $\N=2$ multiplets; 
it was not included in classification of finite theories in section 6.3 of \ci{Fradkin:1985am}
as having  separate $[{3\ov 2}]$  multiplets  that  are not part of an extended conformal  supergravity theory 
is not expected to lead to a classically consistent theory at a non-linear  level.


\section*{{Acknowledgement}}
We  thank S. Kuzenko 
 for useful remarks. We thank C. Closset for  
 comments  and  clarifications regarding 
theories with non-abelian gauge symmetry. 
The work of AAT  was  supported by the STFC grant ST/P000762/1  
and  by  the Russian Science Foundation grant 14-42-00047 at Lebedev Institute.

\appendix

\section{Free action for the $[\tfrac{1}{2}]$ multiplet} 
\la{app:tensor1}

Let us first recall the action  for the  standard chiral multiplet  described  by  a chiral superfield
$\Phi = \phi(y)+\theta\psi(y)+\theta^{2}\,\varphi(y)$  (with  $y_{\alpha\dot \alpha}=
x_{\alpha\dot \alpha}+i\,\theta_{\alpha}\bar \theta_{\dot \alpha} $)
\be
\la{A.2}
S = \int d^{4}x\,d^{4}\theta\,\,\,\Phi^{\dagger}\Phi\to 
\int d^{4}x\,\Big[\phi^{*}\Box\phi+\psi^{\alpha}
\partial_{\alpha\dot \alpha}\overline\psi^{\dot \alpha}+\vp^{*}\vp\Big].
\ee
If instead   one  starts   with  a  chiral spinor superfield 
\be
\la{A.3}
\Phi_{\alpha} = \chi_{\alpha}+\theta^{\beta}\,Q_{\alpha\beta}+\theta^{2}\,\psi_{\alpha},\qquad \qquad Q_{\alpha\beta}=T_{\alpha\beta} + \varepsilon_{\alpha\beta} \phi\ , 
\ee
where $\chi_{\alpha}$ is a spinor with 
dimension $\frac{1}{2}$, $\psi_{\alpha}$ is a standard dimension $\frac{3}{2}$ spinor
and  the boson 
$Q_{\alpha\beta}$ is a 
combination of  symmetric tensor   $T_{\alpha\beta}$ (corresponding to   self-dual part of the antisymmetric tensor $T_{\m\n}$) 
and a  complex scalar $\phi$.
The  corresponding  conformally  invariant action is then
\be
\la{A.5}
S = \int d^{4}x\,d^{4}\theta\,\Phi^{\alpha}\partial_{\alpha\dot \alpha}\overline \Phi^{\dot \alpha} \ .
\ee
In components this gives
\begin{align}
\la{A.6}
S = \int d^{4}x\,\Big[\chi^{\alpha}\,\Box \partial_{\alpha\dot \alpha}\overline\chi^{\dot \alpha}
+Q^{\alpha\beta}\,\partial_{\alpha\dot \alpha}\partial_{\beta\dot \beta}
\,\overline Q^{\dot \alpha\dot \beta}
+\psi^{\alpha}\partial_{\alpha\dot \alpha}\overline \psi^{\dot \alpha}\Big] \ , 
\end{align}
where  the bosonic term  may be written explicitly as 
\be
\la{A.7}
\int d^{4}x\,Q^{\alpha\beta}\,\partial_{\alpha\dot \alpha}\partial_{\beta\dot \beta}\,
\overline Q^{\dot \alpha\dot \beta} = 
\int d^{4}x\,\Big[\phi^{*}\Box \phi+T^{\alpha\beta}
\partial_{\alpha\dot \alpha}\partial_{\beta\dot \beta}\overline T^{\dot \alpha \dot \beta}
\Big].
\ee
Eqs.  (\ref{A.6}),(\ref{A.7}) give   the action for the  superconformal $[\tfrac{1}{2}]$
multiplet in \rf{33.1} (after the renaming $\chi \to \psi^{(3)}$). The  action   for the antisymmetric tensor in \rf{A.7}
in spinor notation  is equivalent to $\del^\m T^+_{\m\n} \del_\l T^{-\l \n}$  in vector notation
with $T^+ \to  T_{\alpha\beta}\sim (1,0)$, $T^- \to  \overline T_{\dot \alpha\dot \beta}\sim (0,1)$.

\section{Chiral anomalies of    conformal gravitino and  
non-gauge
tensor field }
\la{app:chiral}

To compute   anomalies of a higher spin field one needs to couple  it to a gravitational and gauge fields (assuming certain chiral transformation properties), take into account  contribution of ghosts, etc. 
One may use, {\em e.g.}, a perturbative approach, 
computing triangle diagrams    corresponding to the matrix element
of the chiral current between the vacuum and a two-graviton state, or two chiral symmetry gauge fields.

An alternative topological  approach is based on 
 relating the  anomaly of the chiral current to the Atiyah-Singer   index
of a certain elliptic operator mapping fields  to fields of opposite chirality. 
This  approach is somewhat heuristic  and is practically useful 
only if the starting (higher-spin) field theory is consistent.
For a  detailed comparison of the perturbative and topological methods
for standard gravitino see \cite{Nielsen:1984es}.

Here we shall  discuss  the  chiral   gravitational and gauge anomalies of the 
conformal gravitino and the antisymmetric non-gauge
tensor field  $T_{\m\n}$  justifying the values of  their coefficients $\k_1$  and $\k_2$ given in 
\rf{3.4}.\footnote{The standard  antisymmetric tensor gauge  field of rank $2n$ 
with self-dual strength $H=dB$ has gravitation anomaly in $d=4n+2$  dimensions \cite{AlvarezGaume:1983ig}, {\em i.e.}  
not in 4 dimensions where $B_{\mu\nu}$ is dual to a scalar.
}
Their embedding into   conformal supergravity means  that it is possible to 
consistently couple them to gravity and the chiral  gauge field.
Using    topological approach  we shall  assume  
the  existence  
 of a suitable elliptic operator whose index computes the chiral anomaly.
Earlier  results for  the  chiral anomaly coefficients  in  \rf{3.4}  can be found in 
 \cite{Nielsen:1984es,Frampton:1985mb,Romer:1985yg,vanNieuwenhuizen:1985dp}.
 
Following  \cite{Romer:1979bh}, let us  consider a 
a compact 4-manifold $M^4$ and a field  belonging to the general spinor bundle  
$X^{mn}\equiv X^{m,n}$,  
  {\em i.e.} a  tensor $X_{(\a_1...\a_m) (\dot \b_1 ...\dot\b_n)}$ with $m$ symmetric spinor indices
and $n$ symmetric dotted spinor indices.
In particular, the   gravitino $\Psi_\m$  corresponds to $X^{2,1} +  X^{1,2}$ (modulo gauge symmetry) 
while the tensor $T_{\m\n}$   to  $X^{2,0} +  X^{0,2}$ ({\em i.e.} to  the sum of the self-dual 
and antiselfdual parts in spinor notation). 
 The index theorem computes the analytical index of the universal 
chirality-swapping elliptic operator $D_{mn}:\,  X^{mn}
\to X^{nm}$ in terms of topological quantities. 
In 4 dimensions, it reads 
\be
\text{ind}\,D_{mn} = \frac{\text{ch}X^{mn}-\text{ch}X^{nm}}{\text{e}(TM)}
\text{td}(TM\otimes\mathbb C),
\ee
where $TM$ is the tangent bundle, and td, e, ch are the Todd class, the Euler class and the Chern character.
It is understood that one has to extract the part of degree 4 and evaluate it on $M^4$. 
\iffa 
According to 
 \cite{Romer:1979bh}, 
\begin{align}
\la{B.4}
&\text{ind}\,D_{mn} = \frac{1}{6}(m-n)(m+1)(n+1)(m+n+2)  \\& -\frac{(m+1)(n+1)}{720}\,\Big[n(n+2)(3n^2+6n-14)  -m(m+2)(3m^2+6m-14)\Big]\,p_1+\dots,\no
\end{align}
where $p_1 = \frac{1}{8\pi^2}\text{tr}(R\wedge R)$ is the Pontryagin class.
\fi 
 If  the fields transform in a 
non-trivial $U(1)$ gauge bundle $V$, we have to multiply the index by the Chern character 
$\text{ch}(V) = 1-c_2+\dots$. 
The degree 4 terms give the   gravitational and pure gauge contributions
to the divergence of the  corresponding chiral current
 (see, {\em e.g.},  \cite{Romer:1978dy,Romer:1979bh,Nielsen:1978ex,Eguchi:1980jx})
\begin{align}
\mc A_{mn} =  &\text{ind} \big[ D_{mn}\,\text{ch}(V) \big] \Big|_{\text{deg 4}} =\notag \\ 
& -\frac{(m+1)(n+1)}{720}\,\Big[n(n+2)(3n^2+6n-14)-m(m+2)(3m^2+6m-14)\Big]\,p_1\notag \\
& -\frac{1}{6}(m-n)(m+1)(n+1)(m+n+2)\,c_2\ , \la{B.2}
\end{align}
where  $c_2=-{1\ov 8 \pi^2}  \text{Tr}(F\wedge F)$  is the second Chern class  
and $p_1 = \frac{1}{8\pi^2}\text{tr}(R\wedge R)$ is the Pontryagin class. 
For a Weyl fermion we  get  (cf. \rf{2.9},\rf{2.11},\rf{3.4})
\begin{align}
\mc A_{1,0} &= -\tfrac{1}{24}\,p_1-c_2.\la{B.3} 
\end{align}
For  the conformal gravitino (taking into account the ghost subtraction  \ci{Frampton:1985mb})
  and the self-dual 2-tensor or  symmetric bispinor, we  get 
\begin{align}
\mc A_{2,1}-\mc A_{1,0} = \tfrac{5}{6}\,p_1-4\,c_2, \qquad\qquad 
\mc A_{2,0} = \tfrac{1}{3}\,p_1-4\,c_2\ .\la{B.4} 
\end{align}
Thus  in units of the Weyl fermion anomaly  the chiral gravitational anomaly of the conformal 
gravitino  is 
$\frac{5}{6}:(-\frac{1}{24}) = -20$ while its gauge anomaly is $4$.\foot{Let us mention 
for completeness  that  in the case  of  the standard gravitino
the total ghost contribution (taking into account chiralities)
leads  to an extra subtraction $-\mc A_{1,0}$ compared to the conformal gravitino
case, {\em i.e.} 
$\mc A_{2,1}-2\,\mc A_{1,0} = \tfrac{7}{8}\,p_1-3\,c_2$,  so that  the chiral gravitational anomaly
coefficient is -21, while the gauge anomaly is +3 
\cite{Nielsen:1984es,Frampton:1985mb,vanNieuwenhuizen:1985dp}.}

For the self-dual antisymmetric tensor  $T_{\alpha\beta}$ 
the chiral gravitation anomaly is $(-1)\frac{1}{3}:(-\frac{1}{24}) = 8$ while its gauge
anomaly is $-4$ (where we have included a  $-1$ factor due to the different statistics of the tensor
with respect to the Weyl fermion). These values are in agreement with those  given in \rf{3.4}.

The reason why the real $T_{\m\n}$ tensor contributes  \ci{Romer:1985yg}
 to the chiral anomaly can be   understood   from the  analogy of its  kinetic operator 
 in spinor  basis in \rf{A.7}   with the (square of)  Dirac operator for a Weyl spinor. 
Its  chiral gravitational anomaly can be also obtained by adapting  the analysis of \cite{Erdmenger:1999xx}, {\em i.e.} by observing  that  the  antisymmetric tensor 
anomaly  related to  a chiral rotation between the  self-dual and anti-selfdual parts 
 is the same as the electromagnetic duality  anomaly of a  Maxwell field (cf. also \ci{Dolgov:1988qx,Carrasco:2013ypa}).

\section{
Conformal higher spin  partition function on $S^1_\beta\times S^3$}
\la{app:CHS}
Here we shall review the expression for the conformal higher spin partition function  $Z(\b)$ on 
$S^1_\beta\times S^3$   and consider  the   small $\b$ expansion  of $\log Z(\b)$ 
focussing on the interpretation of the  coefficient of the $\log \b$ term. 

Let us  start with the   bosonic fields. 
 The Maxwell vector and the conformal graviton are simplest  cases  of  the 4d conformal higher spin fields 
 with $\Box^s$  kinetic term for a spin $s$ field \cite{Fradkin:1985am}. 
The standard partition function (or character of the corresponding representation of the conformal group) 
for  the  conformal higher spin $s$  field on $S^1_\beta\times S^3$
is given by   \cite{Beccaria:2014jxa} 
\begin{align} 
\la{5.10}
&\log Z_s(\beta) = \sum_{n=1}^\infty\frac{1}{n}\,\mc Z_s(n\b) \ , \ \ \ \ \ 
\\
\la{5.9}
&\mc Z_s(\beta) = \frac{2\,(2s+1)\,t^2-2\,(s+1)^2\,t^{s+2}+2\,s^2\,t^{s+3}}{(1-t)^4},\qquad \qquad t=e^{-\beta}\ . 
\end{align} 
Here $\mc Z_s$ is the single-particle partition function playing the 
role as  the single-particle
index in (\ref{4.1}). 


\def \kkk  {{\rm k}}\def \nn {{\rm n}}

Using the method  sketched in  \rf{4.14}--\rf{4.16}  one  finds that  the  small $\beta$ expansion
of $\log Z_s$  has the form\footnote{Note that on general grounds 
the leading term  in \rf{5.11}   should scale as  $\sim \beta^{-d}$ where $d$ is the space-time dimension.
The reason why the  first term in the
corresponding  expansion of the  supersymmetric partition function in \rf{2.5}  has "softer" 
 $\beta^{-1}$ behaviour  is due to supersymmetric cancellations.}
\begin{align}
\la{5.11}
& \log Z_s =  \frac{\pi^4\,s\,(s+1)}{45\,\beta^3}-\frac{\pi^2\,s(s+1)(s^2+s+1)}{18\,\beta}
+ \kkk_s
-  n_{_\text{CKT}}\log\beta + E_{\rm cas} \,\beta+\mc O(\beta^2) \ , \\
& n_{_\text{CKT}}= \frac{s^2(s+1)^2(2s+1)}{12}\ , \qquad
E_{\rm cas} =\frac{s(s+1)(18s^4+36s^3+4s^2-14s-11)}{720}\,\beta \ . \la {511}
\end{align}
Here 
 the coefficient of the $\log \b$  is simply minus the 
number $n_{_\text{CKT}}$  of conformal conformal Killing tensors in 4 dimensions.\footnote{For a  spin $s$   field  it is the dimension of the $SO(4, 2)$ 
representation $(s - 1, s - 1, 0)$ labelled by the Young tableau that has two rows  of length  $s - 1$.}  $E_\text{cas}$ is the  standard Casimir energy. The constant 
$\kkk_{s}$   in   (\ref{5.11}) 
 has a 
non-polynomial dependence on $s$ and may be expressed as a linear combination of transcendental
constants.\footnote{It can be written as 
\begin{align}
\text{k}_{s} =&-s\,(s+1)\, \frac{\zeta(3)}{4\,\pi^2} +
\frac{1}{6}\,s\,(s+1)\,(s^2+s+1)\,\log(2\pi) -\frac{1}{3} (s+1) (s+2) s^3\, \text{log$\Gamma
   $}_0(s+3)\notag \\
   &+\frac{1}{3} \left(3 s^2+6 s+2\right)
   s^2\, \text{log$\Gamma $}_1(s+3)-(s+1) s^2\,
   \text{log$\Gamma $}_2(s+3)
   +\frac{1}{3} s^2\,
   \text{log$\Gamma $}_3(s+3)\notag \\
   & -\frac{1}{3} (s+1)^2
   \left(3 s^2-1\right)\, \text{log$\Gamma
   $}_1(s+2)+\frac{1}{3} (s-1) (s+1)^3 s\,
   \text{log$\Gamma $}_0(s+2)+(s+1)^2 s\,
   \text{log$\Gamma $}_2(s+2)\notag \\
   &-\frac{1}{3} (s+1)^2\,
   \text{log$\Gamma $}_3(s+2),\no 
\end{align}
where Bendersky's generalised gamma function 
$\text{log$\Gamma $}_k(n) =  \sum_{j=1}^{n-1} j^k\log j$ comes from  simplification of 
derivatives of Hurwitz zeta function by 
$\text{log$\Gamma $}_k(n) = \zeta'_\text{H}(-k,n)-\zeta'(-k)$.
}

Similarly, for the  fermionic conformal   fields with spin 
$s=\s+\frac{1}{2}$ (with $\s=0$  corresponding to  Weyl  spinor, $\s=1$ to  conformal gravitino, etc.) 
one finds  \cite{Beccaria:2016tqy}
\be
\label{5.12}
\mc Z_{\textrm{s}}(\b) = 4\,\frac{(\textrm{s}+1)\,t^{\frac{3}{2}}+(\textrm{s}+1)\,t^{\frac{5}{2}}-(\textrm{s}+1)(\textrm{s}+2)\,t^{\frac{5}{2}+\textrm{s}}
+\textrm{s}(\textrm{s}+1)\,t^{\frac{7}{2}+\textrm{s}}}{(1-t)^{4}} \ .
\ee
The  $\b\to 0$ expansion  of the (periodic)  fermionic  partition function  is given by 
\begin{align}
&\log Z_{\rm s} = -\frac{2\pi^4(\s+1)^2}{45\beta^3}+\frac{\pi^2(\s+1)^2(2\s+1)(2\s+3)}{36\beta}
+\text{k}_\text{s}
+  n_{_\text{CKS}}\log\beta + E_{\rm cas} \,\beta+\mc O(\beta^2) \ , \no \\
&n_{_\text{CKS}} = \frac{\s(\s+1)^3(\s+2)}{3}  , \qquad 
E_{\rm cas} = -\frac{(\s+1)^2(144\s^4+576\s^3+584\s^2+16\s-51)}{2880}. \la{5.13}
\end{align}
Here $   n_{_\text{CKS}}  $ is the number of the conformal Killing spinor-tensors.
As in the bosonic case,  the constant $\text{k}_\text{s}$  has a non-polynomial dependence on $\s$.

\medskip

To further investigate why the coefficient of $\log\beta$ is 
related to the integers  $n_{_\text{CKT}}$ and 
$n_{_\text{CKS}}$ 
let us write  
a general single particle partition functions associated with  a conformal field in 4d as
\be
\la{w.8}
\mc Z(\beta) = \frac{P(t)}{(1-t)^4},\qquad\qquad  P(t) = \sum_q c_q\,t^q,\qquad\qquad t=e^{-\beta},
\ee
where $q$ runs over  some finite set of integers or half-integers. 
This covers the cases
of the  bosonic and fermionic  conformal higher spin fields in  (\ref{5.9}),(\ref{5.12})
 as well as  other "matter" conformal
fields in section 3.1 above  \cite{Beccaria:2014jxa}
\begin{align}
\la{w.8a}
P_\phi(t) &= t(1-t^2), & P_{\pf }(t) &= 1-t^4, &  P_T(t) &= 6\,t(1-t^2),\notag \\
 P_\psi(t) &= 4\, t^{3\ov 2}\,(1-t),  & P_{\psi^{(3)}}(t) &= 4t^{1\ov 2}\,(1-t^3).
\end{align}
For a general $P(t) $ one finds using 
 \rf{5.10}\footnote{This expression is   formally valid for for all fields 
as we assume  that the fermions  are also taken   to be periodic on the "thermal" cycle $S^1_\beta$.}
\begin{align}
\la{w.9}
\log Z(\beta) &=  \frac{\zeta(5)\, p_0}{\beta^4}+\frac{\pi^4\,(2p_0-p_1)}{90\,\beta^3}
+\frac{\,\zeta(3)\, (3p_2-9p_1+11p_0)}{6\,\beta^2} \\
&\ \ -\frac{\pi^2\,(p_3-3p_2+6p_1-6p_0)}{36\,\beta}-\frac{30p_4-60p_3+150p_2-270p_1+251p_0}{720}
\log\beta\notag \\
&\ \ +\text{k}_{P} +\frac{6p_5+10p_3-30p_2+57p_1-54p_0}{1440}\,\beta+\mc O(\beta^2)\ , \qquad 
p_n \equiv   
\left. \frac{d^n}{dt^n}P(t)\right|_{t=1}.\no 
\end{align}
The transcendental
constant $\text{k}_{P}$ is determined by the form of $P(t)$, but cannot be expressed
as a linear combination of its derivatives at $t=1$.
For a theory on $S^1_\beta\times S^3$, the $\beta\to 0$ or $t\to 1$ singularity of $\mc Z(\beta)$
is $\sim1/(1-t)^{3}$, {\it i.e.} $P(t)$ should have an explicit $1-t$ factor.\footnote{This follows both from  the conformal group representation theory
and from the simple
remark that $1/(1-t)^3$ factor  takes into account  the contributions 
of  states on $S^3$   corresponding to all spatial derivatives  of the field.}
This implies the constraint $p_0=P(1)=0$.
Besides,  for all conformal fields we get  one  extra  constraint  on $P(t)$ \footnote{This relation 
was  discussed in \cite{Cherman:2015rpc} where it was related  to the absence of suitable 
counterterms in the heat kernel calculation of the "energy" $E(\b) =-\partial_\beta\log Z(\b)$.
}
\be
\la{w.10}
P''(1)-3\,P'(1)=0 \ ,
\ee
that implies the vanishing of the   coefficient of the $1/\beta^2$ term  in (\ref{w.9}).\
Assuming these constraints, the expansion (\ref{w.9}) simplifies  to
\begin{align}
\la{w.11}
\log Z(\beta) = & -\frac{\pi^4\,p_1}{90\,\beta^3}
-\frac{\pi^2\,(p_3-3p_1)}{36\,\beta}+\text{k}_{P}-\frac{p_4-2p_3+6p_1}{24}\log\beta\no\\
&+\frac{6p_5+10p_3-33p_1}{1440}\,\beta+\mc O(\beta^2) .
\end{align}
This  has similar structure  as the  expansion of the superconformal index 
in (\ref{2.6}) (apart from 
the leading $1/\beta^3$ term that cancels  in a supersymmetric combinations of fields).

The coefficient of the $\log\beta$ term  is thus related to a particular combination 
$p_4-2p_3+6p_1$  of  derivatives of $P(t)$ at $t=1$. To  understand the meaning  of  this combination  we may  use  the  relation between  the 4d  conformal partition function 
and its AdS$_5$  counterpart.\foot{Let us note  that   use  of the AdS  connection is useful but 
is not really necessary  for the final conclusion  as  the form  of $\mc Z(\b)$ used 
below   can be justified purely on the basis of the conformal group representation theory discussed in Appendix F in     \cite{Beccaria:2014jxa}.}
As discussed in   \cite{Beccaria:2014jxa} 
there exists a close relation between the single-particle partition function $\Z$ of a conformal field on 
$S^1_\beta\times S^3$ and the partition function $\mc Z_\text{HS}$ 
of the associated   higher spin field in $\text{AdS}_5$ with quantum numbers determined by the Lorentz spins and 
conformal dimension of the 4d conformal field:\footnote{Changing notation slightly  here instead of $\b$ we use $t=e^{-\beta}$ as the argument of the partition functions.}
\be
\la{w.12}
\mc Z(t)  = \mc Z_\text{HS}(t^{-1})-\mc Z_\text{HS}(t)+\sigma(t) \ . 
\ee
Here the function $\sigma(t)$ is   a finite polynomial in $t+t^{-1}$    and is 
generically present  in the case of  4d conformal  higher   spin fields    related to 
 massless  higher   spin  fields in      $\text{AdS}_5$. It is 
 given by  the character of the finite dimensional irreducible 
representation of  $SO(4,2)$ corresponding, in bosonic  case,  to the conformal Killing tensors in 4 dimensions. 
Its value at $t=1$ gives the dimension of this  representation, {\em i.e.} the total number of conformal
Killing tensors $\sigma(1) = n_{_\text{CKT}}$.\footnote{\la{foot:sigma}The general  expression for $\sigma(t)$ for a bosonic conformal spin $s$ field 
 is

 $\sigma_s(t) = 
\frac{1}{6}\,s\,(s+1)\,(s^{2}+s+1) -\frac{1}{6}\,\sum_{p=1}^{s-1}p\,(p+1)\,\big[ (2s+1)p -3s^{2}-2s-1\big] \,
(t^{s-p}+t^{-s+ p})$.

 Some  special  cases are $\sigma_1(t)=1$, $\ \sigma_2(t)=7+4\,(t+t^{-1})$, 
$\ \sigma_3(t)=26+20\,(t+t^{-1})+9\,(t^2+t^{-2})$.
}
Using  the conformal group representation theory  and \rf{w.12}   one can show that 
\begin{align}
\la{w.13}
\mc Z_\text{HS}= {Q(t)\ov (1-t)^4} \ , \quad {\rm i.e. } \quad  P(t) \equiv  (1-t)^4\,\mc Z(t) 
= 
t^4\,Q(t^{-1})-Q(t)+(1-t)^4\,\sigma(t), 
\end{align}
where   $Q(t)$ is a smooth function. 
The conditions $P(1)=0$ and (\ref{w.10})  can be checked to 
 hold automatically 
and  the  coefficient of the $\log\beta$ term in (\ref{w.11}) then  reads as 
\be
-\frac{p_4-2p_3+6p_1}{24}= -\sigma(1) = -n_{_\textrm{CKT}}\ . \la{c14}
\ee
Thus     the coefficient of  the logarithmic  term is  simply
 minus the number of  the conformal
  Killing tensors  of a rank related to the spin of the conformal gauge  field. Similar result is found for the fermionic  conformal higher spin fields. 
  
 For  non-gauge conformal fields (like the  scalars  $\phi$, $\pf $, 
 spinors $\psi$, $\psi^{(3)}$,  and the tensor $T_{\mu\nu}$)
 one finds  that $\sigma(t)=0$ and thus there is no 
$\log\beta$ term in the small $\b$ expansion of the corresponding  $\log Z(\b)$.

\section{Expansion of partition function   in terms of regularized  theory on $S^3$}
\la{app3d}

It is  possible   to  compute  the constant and logarithmic contributions to 
the small $\beta$ expansion of the partition function directly  in terms of the  spectrum 
of the dimensionally reduced 
3d theory. 
Below  we shall  explain this starting with 
the example of  the standard (non-supersymmetric) partition  function on $S^1_\beta\times S^3$.

\def \wC {{\widehat C}}   \def \rk  {{\rm k}} \def \nn {{\rm n}}

\subsection{Standard  bosonic partition function}

In general, for  a  free conformal field on $S^1_\beta\times S^3$, we can write the single particle partition function  and the   full partition function 
in terms of the  (square roots of)   eigenvalues   $\lambda_n$  and their multiplicities $\text{d}_n$ of the corresponding Laplacian on $S^3$  (equal to energies of states or dimensions of CFT operators) 
\be
\la{dd16}
\Z(\beta) = \sum_n \text{d}_n\,e^{-\b \lambda_n }\ , \qquad \qquad \log Z(\beta) = -\sum_n \text{d}_n\,\log(1-e^{-\beta\lambda_n}).
\ee
Instead of following 
the systematic  derivation of the small $\beta$ expansion of $\log Z$  following   the approach of 
 section \ref{sec:zeta},  one  may attempt  the  direct  $\b \to 0$ expansion of the expression for 
 $\log Z$ in 
(\ref{dd16}):\foot{
The leading singular   $1/\beta^n$ terms  (cf. \rf{w.9}) are   not  explicit 
 in this  naive expansion   approach.} 
\begin{align}
\la{dd18}
&\log Z(\beta) 
{=} -\sum_n \text{d}_n\, \log\beta\, -\sum_n \text{d}_n \log\lambda_n+...
\to   - \nn \,\log \b + \rk + ...\ , \\
& \nn =  \sum_n \text{d}_n\Big|_{\rm reg}\ , \qquad \qquad 
\rk =  -\sum_n \text{d}_n \log\lambda_n \Big|_{\rm reg} \ . \la{dd0}
\end{align}
Thus   the  coefficient of $\log\beta$   should  be  directly related to the (regularized) sum 
of the 
multiplicities, while the constant term $\rk$   should be the  partition function of the reduced 
3d theory on $S^3$.
Note that  as in a  conformal theory $Z$ depends only on dimensionless ratio $\b/R$ where $R$ is the radius of $S^3$  (the square roots of eigenvalues $\l_n$ scale as  $R^{-1}$)
 the dependence on $R$ is also   controlled by  the regularized total number of the eigenvalues   or $\nn$.\footnote{Notice that $\nn$ is  also the coefficient of the 
 $\beta^{0}$ term in the small $\beta$ expansion of the single particle partition function 
 $\Z(\beta)$, cf. (\ref{dd16}).}

 The natural regularization  is the spectral $\z$-function  one:
 if 
 $\z_\Delta(z)= \sum_n  \text{d}_n \l_n^{-z}$ then $\nn=\z_\Delta (0)$ and  $\rk=\zeta_\Delta'(0$).  This analytic  regularization can be implemented simply by 
 adding the  factors $e^{-\eps \l_n}$, doing the sums and then dropping all  terms   which are singular in the limit $\eps\to 0$. 

As an example,  let us  consider the  partition function of a  conformally coupled
scalar  on $S^1_\beta\times S^3$  where  ($t\equiv  e^{-\b}$)
\be
\la{dd19}
\mc Z_0(\beta) = \frac{t-t^3}{(1-t)^4} = \sum_{n=1}^\infty n^2\,t^n,\qquad \qquad 
\log Z_0(\beta) = -\sum_{n=1}^\infty n^2\,\log(1-e^{-\beta\,n}).
\ee
Using the  method  of  section \ref{sec:zeta}  one can show that 
 the  exact small $\beta$ expansion of $\log Z_0$  is 
\be
\la{dd20}
\log Z_0(\beta) = \frac{\pi^4}{45\,\beta^3}
+\frac{0}{\beta}+0\cdot\log\beta-\frac{\zeta(3)}{4\pi^2}+\frac{1}{240}\b+\mc O(e^{-1/\beta}).
\ee
The same   results of the  coefficient $\nn=0$ of $\log \b$  term and 
the constant  term $\rk= -\frac{\zeta(3)}{4\pi^2}$ are indeed   found by using 
directly the  corresponding regularized expressions in \rf{dd0}
(here $\l_n = n$   and ${\rm d}_n=n^2$)
\begin{align}
&\nn= \sum_{n=1}^\infty e^{-\eps\,n}\,n^2 \Big|_{\eps^0}
 = \frac{2}{\eps^3}+\mc O(\eps) \Big|_{\eps^0}\  = \  0 \ , \\
& \rk=-\sum_{n=1}^\infty e^{-\eps\,n}\,n^2\,\log n  \Big|_{\eps^0}= 
\frac{2\log\eps}{\eps^3}+\frac{-3+2\,\gamma_\text{E}}{\eps^3} -\frac{\zeta(3)}{4\pi^2}  
 +\mc O(\eps)\ \Big|_{\eps^0}= \ -\frac{\zeta(3)}{4\pi^2}\ . \la{dd77}
\end{align}
Thus  the constant in (\ref{dd20})  may be identified with the  partition function
of the dimensionally reduced scalar 3d theory on $S^3$ computed using natural analytic regularization.\footnote{Note that 
  the dimensionally reduced 3d theory   does not,   of course,  
   correspond to a conformal scalar on $S^3$: the  4d conformal  scalar operator    $-\nabla^2+\frac{R}{6}$ 
  reduces to the same one on  $S^3$  (with $R$  here being the curvature of $S^3$) 
    while   the  conformally 
  coupled scalar on $S^3$ would have  the   kinetic operator 
 $-\nabla^2+\frac{R}{8}$.
}

Similar  computation    can  be done  for the Maxwell  vector field where 
\be\la{dd88}
\mc Z_{1}(\beta) = \sum_{n=1}^\infty 2n(n+2)\,t^{n+1},\qquad \quad
\log Z_1(\beta) = -\sum_{n=1}^\infty 2\,n(n+2)\,\log(1-e^{-\beta\,(n+1)}).
\ee
Here $\l_n = n+1$   ($n=1,2, ....$) is the square root of the eigenvalue of the transverse 3-vector Laplacian on $S^3$  and $2n(n+2)$  is its degeneracy  \cite{Beccaria:2014jxa}. 
The exact  small $\beta$ expansion of $Z_1$   computed  as in section \ref{sec:zeta}  reads 
\be
\la{dd24}
\log Z_1(\beta) = \frac{2\pi^4}{45\,\beta^3}
-\frac{\pi^2}{3\beta}-\log\beta+\log(2\pi)-\frac{\zeta(3)}{2\pi^2}
+\frac{11}{120}\b+\mc O(e^{-1/\beta}).
\ee
Using instead  the direct expansion and   \rf{dd0}  we find
  gives 
\begin{align}
& \nn=  \sum_{n=1}^\infty e^{-\eps\,(n+1)}\,2\,n(n+2)\Big|_{\eps^0} = 
\frac{2(3 e^{\epsilon }-1)}{\left(e^{\epsilon
   }-1\right)^3} \Big|_{\eps^0} = \Big[\frac{4}{\eps^3}-\frac{2}{\eps}+1+\mc O(\eps)\Big]_{\eps^0} = 1 ,\no \\
&\rk= -\sum_{n=1}^\infty e^{-\eps\,(n+1)}\,2\,n(n+2)\,\log(n+1) \Big|_{\eps^0}= 
\frac{4 \log \eps+4 \gamma_\text{E} -6}{\epsilon
   ^3}+\frac{-2 \log \eps-2 \gamma_\text{E}
   }{\epsilon }\notag \\
   &\qquad \qquad \qquad \qquad\qquad \qquad+\log (2\pi)-\frac{\zeta (3)}{2 \pi
   ^2}+\mc O(\eps)\Big|_{\eps^0} =  \log (2\pi)-\frac{\zeta (3)}{2 \pi
   ^2} \ ,  \la{dd11}
\end{align}
in agreement with the  coefficients of the  $\log\beta$  and the constant term in (\ref{dd24}).

Another 
 non-trivial example   is that of the  conformal graviton  for which \cite{Beccaria:2014jxa}\foot{Here  to determine the effective $\l_n = n+2, \ {\rm d}_n = 2\,(3\,n^2+12n+5)$
 we re-expanded   the  final expression for the   single-particle partition function 
 of the conformal graviton on $S^1 \times S^3$   in Eq. (3.22)  of \cite{Beccaria:2014jxa}
 that was obtained by combining the contributions of the transverse  graviton  and vector 
 Laplacians on $S^3$.  In the notation of \cite{Beccaria:2014jxa}, these are respectively
 $\Z_{2,0} = \sum_{n=0}^{\infty}2\,(n+1)(n+5)\,(t^{n+2}+t^{n+4})$ and
 $\Z_{1,1} = \sum_{n=1}^{\infty}2\,(n+1)(n+3)\,t^{n+2}$ with $\Z_{2}=\Z_{2,0}+\Z_{1,1}$.
 }
\begin{align}
\Z_2(\beta) = &\sum_{n=0}^\infty 2\,(3\,n^2+12n+5)\,t^{n+2} \ , \\
\nn=& 
\sum_{n=0}^\infty e^{-\eps(n+2)} 2\,(3\,n^2+12n+5) 
\Big|_{\eps^0}
=  \frac{2 (9 \sinh \eps+\cosh\eps+5)}{\left(e^{\eps}-1\right)^3}
\Big|_{\eps^0}\no \\  = &
\frac{12}{\eps^3}-\frac{14}{\eps
   }+15+\mc O(\eps)
\Big|_{\eps^0} = 15 \ , \la{dd12}
\end{align}
in agreement with (\ref{5.14}).

Similar computations  can be  done  also for  other conformal fields
appearing in the $\mc N=1$ multiplets  discussed in the text,  confirming that
 $\nn=0$ for non-gauge fields  and is  always an    integer (see \rf{513},\rf{511},\rf{5.13},\rf{c14})
for the  gauge fields. 
This fact   suggests that it should   have some "zero-mode" interpretation 
which remains to be  clarified
(cf. \ci{Giombi:2013yva}). 

\subsection{Supersymmetric case}
\la{app3d-susy}

The supersymmetric  partition  function  is the  same as  the superconformal index up 
 to the normal ordering  supersymmetric Casimir  energy  factor  in (\ref{2.2}). This means that we may 
use the expansion of the index to extract the analogs of $\text{d}_n$ and $\lambda_n$ in (\ref{dd16}).
These will have  again the  meaning of multiplicities and eigenvalues of the single particle 
(free) supersymmetric spectrum. 

Application of (\ref{dd0})  is expected to give the constant and logarithmic
terms in the expansion of the index. Let us check this claim for   few examples  of 
 $\mc N=1$ multiplets  using the expressions in Table \ref{tab:indices}.
For the chiral multiplet, we have 
\be
\text{i}_{[0]}(\beta) =\frac{t^\frac{2}{3}-t^\frac{4}{3}}{(1-t)^2} = 
\sum_{n=0}^\infty (n+1)\,(t^{n+\frac{2}{3}}-t^{n+\frac{4}{3}}).
\ee
Hence, from (\ref{dd0}),
\begin{align}
&\nn_{[0]}= \sum_{n=0}^\infty \Big[e^{-\eps\,(n+\frac{2}{3})}-e^{-\eps\,(n+\frac{4}{3})}\Big]\,(n+1) \Big|_{\eps^0}
 = \frac{2}{3\,\eps}+\mc O(\eps) \Big|_{\eps^0}\  = \  0 \ , \\
& \rk_{[0]}=-\sum_{n=0}^\infty \Big[e^{-\eps\,(n+\frac{2}{3})}\log(n+\tfrac{2}{3})
-e^{-\eps\,(n+\frac{4}{3})}\log(n+\tfrac{4}{3})\Big]\,(n+1) \Big|_{\eps^0}\notag \\
&\ \ \ \ =  \lim_{a\to 0}\partial_a\Bigg\{
\tfrac{1}{3} e^{-\frac{4}{3} \epsilon} \Big[
3 \Phi (e^{-\epsilon},-a-1,\tfrac{4}{3})-\Phi(e^{-\epsilon },-a,\tfrac{4}{3})\Big]\notag \\
&   -\tfrac{1}{3}\,e^{-\frac{2}{3} \epsilon} \Big[3 \Phi(e^{-\epsilon },-a-1,\tfrac{2}{3})
   +\Phi(e^{-\epsilon },-a,\tfrac{2}{3})\Big]
 \Bigg\}\Bigg|_{\eps^0}\  = \    \frac{\pi }{9 \sqrt{3}}-\frac{1}{6} \log
   3 -\frac{\psi ^{(1)}\left(\frac{1}{3}\right)}{6
   \sqrt{3} \pi }\ , 
\end{align}
where $\Phi(z, s, \alpha)=\sum_{k=0}^\infty\frac{z^k}{(k+\alpha)^s}$ 
is the Lerch function and we expanded around $\eps=0$ using 
\be
\Phi(z, s, \alpha) = \zeta(s,\alpha)+(z-1)\,[\zeta(s-1, \alpha+1)-\alpha\zeta(s, \alpha+1)]+\dots.
\ee
As a result, these values  of $\nn_{[0]}=0$ and $\rk_{[0]}$ are    in agreement with (\ref{4.21}).

For the vector multiplet
\be
\text{i}_{[1]}(\beta) =\frac{-2t+2t^2}{(1-t)^2} = -2\,\sum_{n=1}^\infty t^n,
\ee
 and  applying again (\ref{dd0}) we find 
\begin{align}
&\nn_{[1]}= -2\,\sum_{n=1}^\infty e^{-\eps\,n}\Big|_{\eps^0}
 = -\frac{2}{\eps}+1+\mc O(\eps) \Big|_{\eps^0}\  = \  1 \ , \\
& \rk_{[1]}= 2\,\sum_{n=1}^\infty e^{-\eps\,n}\,\log n \Big|_{\eps^0}
 =  \frac{-2\,\log\eps-2\,\gamma_\text{E}}{\eps}+\log(2\pi)+\mc O(\eps)\Big|_{\eps^0}\  = \  
 \log(2\pi),
 \end{align}
 in agreement with  (\ref{4.21}).
  
  In the case of the  graviton   multiplet  $[2]$ we  have  
 \be
\text{i}_{[2]}(\beta) =\frac{-4t+4t^3}{(1-t)^2} = -4t-8\,\sum_{n=2}^\infty t^n,
\ee
and  then  the values of $\nn_{[2]}$, $\rk_{[2]}$ are, again, in agreement   with  (\ref{4.21})
\begin{align}
&\nn_{[2]}= -4-8\,\sum_{n=2}^\infty e^{-\eps\,n}\Big|_{\eps^0}
 = -\frac{8}{\eps}+8+\mc O(\eps) \Big|_{\eps^0}\  = \  8 \ , \\
& \rk_{[2]}=8\,\sum_{n=2}^\infty e^{-\eps\,n}\,\log n \Big|_{\eps^0}
 =  4\,\text{k}_{[1]} = 4\,\log(2\pi)\ .
 \end{align}
 Similar  agreement with values in  (\ref{4.21})  is  found also for other  multiplets.

\def \rz {{\rm z}}

\section{Superconformal index  corresponding to   $\mc N=1$ multiplets on  squashed $S^3$}
\la{app:squashed}

In  this Appendix we shall   consider the generalized  2-parameter  superconformal 
  index \rf{2.1},\rf{4.1} 
which happens to be related   \cite{Imamura:2011wg,Closset:2012ru,Aharony:2013dha} to the supersymmetric partition 
on $S^1_\beta\times S^3_b$    where $b$ is the squashing parameter of the 3-sphere.
The corresponding choice of   the fugacities   generalizing $p=q=e^{-\b}$ in \rf{00},\rf{2.4} is
 \be p=e^{-\beta/b}\ , \qquad \qquad 
q=e^{-\beta\,b}\ . \la{d0} \ee
The  small $\b$ expansion of the corresponding   index for the scalar chiral multiplet $[0]$ was  found in 
\cite{Ardehali:2015hya} 
\be
\la{.1}
\log\text{I}_{[0]}(\beta, b) = \frac{b+b^{-1}}{18}\frac{\pi^2}{\beta}+
\Big(\frac{b+b^{-1}}{216}+\frac{b^3+b^{-3}}{162}\Big)\,\beta+\dots.
\ee
The supersymmetric Casimir energy, entering the general relation (\ref{2.2}),
is expected to have the general form 
\cite{Closset:2013sxa,Assel:2014paa}
\be
\la{.2}
E_\tsusy = \frac{2}{9}\,(b+b^{-1})\,\text{a}-\frac{2}{27}\,(b^3+b^{-3})\,(2\,\text{a}-3\,\text{c}),
\ee
that reduces to (\ref{2.3}) for $b\to 1$.
We can easily obtain the expansion (\ref{.1}) by the $\zeta$-function methods  
described  in  Section \ref{sec:zeta}.  Let us first  recall that  for general $p,q$ the single-particle superconformal index in \rf{4.1} 
is \cite{Dolan:2008qi} (reducing to  \rf{4.2} for $b=1$ or $p=q=t$))
\be
\la{.3}
\text{i}_{[0]}(p,q) = \frac{(pq)^\frac{1}{3}-(pq)^\frac{2}{3}}{(1-p)(1-q)}.
\ee
The  Mellin  transform \rf{4.13}  of the index  gets the following  contribution from a term of the form 
$\frac{e^{-a\beta}}{(1-e^{-\beta\,b})(1-e^{-\beta\,b^{-1}})}$ 
\be
\la{.4}
\zeta_2 (u; b, b^{-1}, a) = \sum_{n,m=0}^\infty (a+b\,n+b^{-1}\,m)^{-u} \ , 
\ee
where we adopted  the standard notation for the Barnes double zeta function \cite{ruijsenaars2000barnes}.
Expanding around $u=1,0,-1$ we have  (cf. (\ref{4.14}))
\be
\la{.5}
\beta^{-u}\Gamma(u)\zeta(u+1) = \begin{cases}
\frac{\pi^2}{6\,\beta}+\dots\ , & u\to 1, \\
\frac{1}{u^2}-\frac{1}{u}\log\beta+\dots\ , & u\to 0,\\
\frac{\beta}{2\,(u+1)}+\dots\ , & u\to -1.
\end{cases}
\ee
Using \cite{ruijsenaars2000barnes,spreafico2009barnes}
\begin{align}
\la{.6}
\mathop{\text{Res}}_{u=1} \zeta_2(u; b, b^{-1}, a) &= \frac{b+b^{-1}}{2}-a, \qquad 
\zeta_2(0; b, b^{-1}, a) = \frac{1}{4}+\frac{b^2+b^{-2}}{12}-\frac{a}{2}(b+b^{-1})+\frac{a^2}{2},\notag \\
\zeta_2(-1; b, b^{-1}, a) &= -\frac{b+b^{-1}}{24}+\left(\frac{1}{4}+\frac{b^2+b^{-2}}{12}\right)\,a
-\frac{b+b^{-1}}{4}\,a^2+\frac{a^3}{6}\ , 
\end{align}
we obtain  from (\ref{.3}) the following expression for 
 $\ZZ_\text{m}(u)$ defined in (\ref{4.13}) 
\begin{align}
\la{.7} 
\ZZ_{[0]}(u) &= \zeta_2\big(u; b, b^{-1}, \tfrac{1}{3}(b+b^{-1})\big)-
\zeta_2\big(u; b, b^{-1}, \tfrac{2}{3}(b+b^{-1})\big)\ . 
\end{align}
Combining the results in (\ref{.6}) with (\ref{.5}) we reproduce  (\ref{.1}). 

A similar computation   can be done  for 
 the vector multiplet index where 
\be
\la{d1}   
\text{i}_{[1]}(p,q) = -\frac{p}{1-p}-\frac{q}{1-q} = \frac{
-p-q+2pq
}{(1-p)(1-q)}\ .
\ee
In this case
\begin{align}
\la{d2} 
\ZZ_{[1]}(u) &= -\zeta_2\left(u; b, b^{-1}, b\right)
-\zeta_2\left(u; b, b^{-1}, b^{-1}\right)+2\,\zeta_2\left(u; b, b^{-1}, b+b^{-1}\right),
\end{align}
and using again (\ref{.6}) with (\ref{.5}) we find
\be
\la{d3}
\log\text{I}_{[1]}(\beta, b) = -\frac{b+b^{-1}}{6}\frac{\pi^2}{\beta}+\text{k}([1], b)
-\log\beta+\frac{b+b^{-1}}{24}\,\beta+\dots,
\ee
which  matches the expression in Eq.~(A.19) in \cite{Ardehali:2015hya}. 
Notice that the -1  coefficient of $\log\beta$ is 
independent of $b$, {\em i.e.}  is  the same as in \rf{4.19},  supporting
 its "topological" interpretation.   

The same  analysis  can be  repeated for  the non-unitary multiplets.
For the higher derivative multiplet $[0']$  
 using the data  in  Table \ref{tab:0}  to  sum $(-1)^F p^{j_1+j_2+r/2}\,q^{-j_1+j_2+r/2}$  and 
dividing by $(1-p)(1-q)$ we find 
\be
\la{.8}
\text{i}_{[0']}(p,q) = \frac{1-pq}{(1-p)(1-q)} \ , 
\ee
leading to the   small $\beta$ expansion 
\be
\la{.9}
\log\text{I}_{[0']}(\beta, b) = \frac{b+b^{-1}}{6}\frac{\pi^2}{\beta}+\text{k}([0'], b)
+0\cdot\log\beta-\frac{b+b^{-1}}{24}\,\beta+\dots.
\ee
For the   $[\frac{1}{2}]$ multiplet  in Table \ref{tab:tensor} 
 we obtain   
\be
\la{.10}
\text{i}_{[\frac{1}{2}]}(p,q) = \frac{-p^\frac{2}{3}q^{-\frac{1}{3}}+
p^\frac{4}{3}q^\frac{1}{3}-q^\frac{2}{3}p^{
-\frac{1}{3}}+p^\frac{1}{3}q^\frac{4}{3}}{(1-p)(1-q)}\ ,
\ee
\be
\la{.11}
\log\text{I}_{[\frac{1}{2}]}(\beta, b) = -\frac{2}{9}(b+b^{-1})\,\frac{\pi^2}{\beta}+
\text{k}([\tfrac{1}{2}], b)
+0\cdot\log\beta+\Big[-\frac{2}{27}(b+b^{-1})+\frac{11}{162}(b^3+b^{-3})\Big]\,\beta+
\dots.
\ee
For the gravitino multiplet $[\frac{3}{2}]$   in Table \ref{tab:gravitino} we get  
\begin{align}
\la{.12}
\text{i}_{[\frac{3}{2}]}(p,q) = & \frac{1}{(1-p)(1-q)}\, \Big[
-2 p^{5\ov 3} q^{2\ov 3}-2 p^{4\ov 3} q^{4\ov 3}+p^{4\ov 3} q^{-{2\ov 3}}\notag \\
&\qquad \qquad\qquad\ \    -2
   p^{2\ov 3} q^{5\ov 3}+p^{2\ov 3}
   q^{2\ov 3}+q^{4\ov 3} p^{-{2\ov 3}}+p^{5\ov 3} q^{-\frac{1}{3}}
   +q^{5\ov 3} p^{-\frac{1}{3}}+p^\frac{1}{3}q^\frac{1}{3}\Big],
\end{align}
\be
\la{.13}
\log\text{I}_{[\frac{3}{2}]}(\beta, b) = \frac{13}{18}(b+b^{-1})\,\frac{\pi^2}{\beta}+
\text{k}([\tfrac{3}{2}], b)
+6\,\log\beta+\Big[-\frac{71}{216}(b+b^{-1})-\frac{22}{81}(b^3+b^{-3})\Big]\,\beta+
\dots.
\ee
A   similar result is found for the graviton
multiplet $[2]$ in Table \ref{tab:graviton}  
\begin{align}
\la{.14}
\text{i}_{[2]}(p,q) &= \frac{-p-q-p^2q^{-1}-q^2 p^{-1}+2p^2 q+2pq^2}{(1-p)(1-q)}\ , 
\end{align}
\be
\la{.15}
\log\text{I}_{[2]}(\beta, b) = -\frac{2}{3}(b+b^{-1})\,\frac{\pi^2}{\beta}+
\text{k}([2], b)
+8\,\log\beta+\Big[\frac{2}{3}(b+b^{-1})+\frac{1}{2}(b^3+b^{-3})\Big]\,\beta+
\dots.
\ee
The pattern is thus  the same as in the previous cases. In particular, the coefficient of  $\log\beta$ 
does not depend on $b$ and has the same (integer) value  as 
we found  in the undeformed case (see \rf{4.21}). 

\def \Tr {{\rm Tr\,}} 

\section{Small $\beta$ expansion of  superconformal index for $(1,0)$ multiplets\\ in six dimensions}
\la{app:6d}

The above  discussion of the index  in four dimensions can be readily extended 
 to six dimensional $(1,0)$ superconformal theories. 
Here we   shall  briefly outline what one finds for the unitary scalar and tensor multiplets, as well as for a non-unitary higher derivative vector multiplet. 

The superconformal index for a $(1,0)$   6d  theory is defined similarly to (\ref{2.1}) 
\cite{Bhattacharya:2008zy} 
\be
\text{I}(t, u, v) = \mathop{\text{Tr}}
\Big[(-1)^F\,t^{\Delta-\frac{r}{2}}\,u^{j_1}\,v^{j_2}\Big]_{\Delta = 2\,r+\frac{1}{2}(j_1+2 j_2+3j_3)} 
\ee
Here  $(\Delta, r, j_1, j_2, j_3)$ 
are associated to the subgroups of 
$\text{OSp}(8^*|2)\subset SO(2,6)\times SU(2)_r\supset U(1)_\Delta\times SU(4)\times SU(2)_r$
with  $(j_1, j_2, j_3)$ being  the Dynkin labels of $SU(4)$.
We are interested in the specialization corresponding to a  supersymmetric partition function 
 on $S^1_\beta\times S^5$
\be
\text{I}(\beta) \equiv  \text{I}(e^{-\beta}, 1, 1) \ .
\ee
Leading  terms  in the  $\beta\to 0$ expansion 
 are  expected to be related to the coefficients of the 
8-form polynomial $\mc A_8$ encoding the chiral (R-symmetry and gravitational) anomaly
(which  in turn  are related to the 6d conformal anomaly coefficients). 
 These will play a role
similar to   that of the   $\Tr \RR $ and $\Tr \RR^3$  in \rf{2.10} or to the 
 conformal anomaly coefficients a  and  c in four dimensions. 
The structure of $\mc A_8$
is 
\cite{Frampton:1983ah,AlvarezGaume:1983ig,Zumino:1983rz}
\begin{align}
\mc A_8 &= \tfrac{1}{4!}\,(\alpha\,c_2^2+\beta\,c_2\,\text{p}_1+\gamma\,\text{p}_1^2+\delta\,\text{p}_2),
\notag \\
c_1 = \text{tr} F, \qquad 
c_2 &= \text{tr} F^2, \qquad \text{p}_1 = -\tfrac{1}{2}\,\text{tr} R^2, \qquad \text{p}_2 = -\tfrac{1}{4}\text{tr}R^4+\tfrac{1}{8}(\text{tr} R^2)^2\ . 
\end{align}
According to \cite{DiPietro:2014bca,Liu:2018eml}, the small $\b$ expansion  should have the following  structure 
\be
\la{six-1.3}
\log\text{I}(\beta) = \frac{8\pi^4}{9\,\beta^3}\Big(\gamma+\frac{1}{4}\delta\Big)
+\frac{\pi^2}{6\beta}\Big(\frac{9}{2}\beta-8\gamma+\delta\Big)-
\text{n}\,\log\beta
+ \text{k}
+ E_\tsusy\, \beta +\dots,
\ee
where the six-dimensional supersymmetric Casimir energy is 
 \cite{Bobev:2015kza,Yankielowicz:2017xkf}
\be\la{f5}
E_\tsusy = -\frac{27}{128}\,\alpha+\frac{9}{32}\,\beta-\frac{3}{8}\,\gamma-\frac{1}{8}\delta\ .
\ee
The constant $\text{k}$ and  $\log\beta$ term were  not, in fact,  discussed in 
\cite{DiPietro:2014bca,Liu:2018eml}  (the values of n and k    given below  will thus be new)
 but they   should be  expected from the general analysis
we gave  above  in four dimension.

To test the validity of (\ref{six-1.3}) 
let us  consider the standard  unitary (1,0) scalar $\text{S}^{(1,0)}$ and tensor  $\text{T}^{(1,0)}$ 
6d multiplets and also a non-unitary higher derivative 
multiplet $\text{V}^{(1,0)}$; their  field content and anomaly coefficients are summarized in Table~\ref{tab:six}
(where we indicated  chiralities of the fields). 
The  scalar and tensor multiplets contain  combinations of the  2-derivative scalar $\phi$, the
Majorana-Weyl (MW) spinor $\psi$ and the standard 
  gauge (anti)selfdual 2-tensor $B^-$.
The higher-derivative  (1,0) vector multiplet  $\text{V}^{(1,0)}$ contains 
a 4-derivative  gauge vector $V^{(4)}$ and a 3-derivative spinor $\psi^{(3)}$ (in addition to scalars)  
\cite{Ivanov:2005kz,Beccaria:2015uta}.\foot{This multiplet may be identified with the $n=2$ case of the 
$\mc O^*(n)$ multiplets recently discussed in \cite{Kuzenko:2017jdy}.}
\begin{table}
\setlength\extrarowheight{8pt}
\begin{center}
\begin{tabular}{lcccc}
\toprule
& $\alpha$ & $\beta$ & $\gamma$ & $\delta$ \\
\midrule
$\text{S}^{(1,0)} = 4\,\phi+2\,\psi^-$ & 0 & 0 & $\frac{7}{240}$ & $-\frac{1}{60}$\\
$\text{T}^{(1,0)} = \phi+2\,\psi^-+B^-$ & 1 & $\frac{1}{2}$ & $\frac{23}{240}$ & $-\frac{29}{60}$\\
$\text{V}^{(1,0)} = 3\phi+2\,\psi^{(3)+}+V^{(4)}$ & $-1$ & $-\frac{1}{2}$ & $-\frac{7}{240}$ & 
$\frac{1}{60}$ \\
\bottomrule
\end{tabular}
\end{center}
\caption{Anomaly coefficients of   $(1,0)$ superconformal multiplets in six dimensions.}
\la{tab:six} 
\end{table}

For the 6d scalar and tensor  multiplets the  single-particle superconformal index  is given by   \cite{Buican:2016hpb}
\be
\text{i}_\text{S}(\beta) = 2\,\frac{t^\frac{3}{2}-t^\frac{5}{2}}{(1-t)^4},\qquad\qquad
\text{i}_\text{T}(\beta) = \frac{-3 t^2+4 t^3-t^4}{(1-t)^4}.
\ee
Using the method of Section \ref{sec:zeta}  used  above in 4d  case  we find  
\begin{align}
\log \text{I}_\text{S}(\beta) &= \frac{\pi^4}{45}\,\frac{1}{\beta^3}-\frac{\pi^2}{24}\,\frac{1}{\beta}
+0\cdot \log\beta+\text{k}_\text{S}-\frac{17}{1920}\,\beta+\mc O(e^{-1/\beta}),\\
\log \text{I}_\text{T}(\beta) &= -\frac{\pi^4}{45}\,\frac{1}{\beta^3}+\frac{\pi^2}{6}\,\frac{1}{\beta}
+\frac{1}{2}\log\beta+\text{k}_\text{T}-\frac{11}{240}\,\beta+\mc O(e^{-1/\beta}).\la{f8}
\end{align}
The singular terms $\sim \beta^{-3}$ and $\sim \beta^{-1}$ as well as the Casimir term 
are in full agreement with (\ref{six-1.3}),\rf{f5}. In addition, we thus  find that 
\begin{align}
\text{n}_\text{S} = 0,  \quad  \text{k}_\text{S} = \frac{1}{8}\log 2+\frac{3\zeta(3)}{16\pi^2} \ ,\qquad   \qquad 
\text{n}_\text{T} = -\frac{1}{2}  , \quad  \text{k}_\text{T} = -\frac{1}{2}\log(2\pi)+\frac{\zeta(3)}{4\pi^2} \ .
\end{align}
For the non-unitary $\text{V}^{(1,0)}$ multiplet,  the following educated Ansatz for the 
single-particle index
\be\la{f10}
\text{i}_\text{V}(\beta) = \frac{-3t+6t^2-5t^3+2t^4}{(1-t)^4},
\ee
gives the  following expansion
\be
\log \text{I}_\text{V}(\beta) = -\frac{\pi^4}{45}\,\frac{1}{\beta^3}-\frac{\pi^2}{3}\,\frac{1}{\beta}
-\log\beta+\text{k}_\text{V}+\frac{19}{240}\,\beta+\mc O(e^{-1/\beta}).
\ee
A check of  the  consistency of \rf{f10} is  that 
 the coefficients of the 
  leading   $\sim \beta^{-3}$, $\sim \beta^{-1}$  and $\b$  terms  
are  again in agreement with (\ref{six-1.3}),\rf{f5} and values in  Table \ref{tab:six}.
In addition, we find that 
\be
\text{n}_\text{V} = 1,\qquad \qquad 
\text{k}_\text{V} = \log(2\pi)+\frac{\zeta(3)}{4\pi^2}.
\ee
The constant term $\text{k}$ has a natural interpretation of  the logarithm of the 
partition function of the 5-dimensional reduced theory found in the limit  $\beta\to  0$. 

As for the  $\log\beta$ term, it is 
 present for the tensor and non-unitary vector 
  multiplets. To understand its 
origin,  let us  briefly explain how to  generalize  the 4d analysis of  Appendix \ref{app:CHS}  to the present 6d case. 
As in the  4d  case, we expect that the $\log\beta$ correction  should come  only 
from gauge fields 
in the multiplets. 
These are  the  gauge tensor $B^-$ in the tensor multiplet and 
the 4-derivative gauge  vector $V^{(4)}$  (with  kinetic term $(\del_\mu F^{\m\n})^2$)
in the  vector multiplet. 
We also expect it to be related to the 6d analog of the 
$\sigma(t)$ term in (\ref{w.12}),\rf{c14}, {\em i.e.}  $\text{n}=\sigma(1)$. 

To check this claim,
let us begin with the $V^{(4)}$ field. The associated 
dual higher spin field in AdS$_7$ (see discussion before (\ref{w.12}))  has the $SO(2,6)$ representation content
\cite{Beccaria:2015uta}
\be
\la{six-2.8}
(5; 1,0,0)-(6; 0,0,0),
\ee
where $(\Delta^+; \bm{h})\equiv(\Delta^+; h_1, h_2, h_3)$ denote  $SO(2,6)$ quantum numbers
(here $\Delta^+= 6-\Delta$  where   $\Delta$ is canonical dimension of  6d field).
Each  representation contributes   to the analog of the   partition function in  \rf{w.12} 
as $\mc Z^+_\text{HS}(t) = \text{d}(\bm{h})\,\frac{t^{\Delta^+}}{(1-t)^6}$
where 
\be
\text{d}(\bm{h}) = \tfrac{1}{12}(1+h_1-h_2)(1+h_2-h_3)(1+h_2+h_3)(2+h_1-h_3)(2+h_1+h_3)
(3+h_1+h_2)
\ee
is the dimension of the $SO(6)$ representation with Dynkin labels $\bm{h}$. Thus, the higher spin 
partition function of the AdS$_7$ field transforming as  (\ref{six-2.8})
is given by 
\be
\mc Z^+_\text{HS}(t) = \frac{\text{d}(1,0,0)\,t^5-\text{d}(0,0,0)\,t^6}{(1-t)^6} = \frac{6t^5-t^6}{(1-t)^6}\ . 
\ee
The analog of the representation 
 (\ref{w.12})  then   implies that  the 6d partition function of $V^{(4)}$  is given by 
\begin{align}
\la{six-2.1}
\mc Z_{V^{(4)}}(t) &= \mc Z^+_\text{HS}(t^{-1})-\mc Z^+_\text{HS}(t)+\sigma(t)
=\frac{-1+6t-6t^5+t^6}{(1-t)^6}+\sigma(t) \notag \\
&= (-1+15\,t^2+70\,t^3+\dots)+\sigma(t).
\end{align}
The  $15\,t^2$ term is associated with the lowest dimension field which is 
the field  strength $F_{\mu\nu}$ (which has   indeed dimension 2
and $\binom{6}{2}=15$ components). Hence, we need 
$\sigma(t)=+1$ in order to cancel the spurious $-1$  and get  the correct 6d partition function
for the conformal field $V^{(4)}$.\footnote{The pattern is completely similar to the case of a Maxwell vector in 4d  where the dual 
higher spin partition function in AdS$_5$ is the $s=1$ case of 
$\mc Z^+_{\text{HS}, s}(t) = \frac{(s+1)^2\,t^{s+2}-s^2\,t^{s+3}}{(1-t)^4}$ \cite{Beccaria:2014jxa}. 
This gives
$\mc Z_1(t) = \mc Z^+_{\text{HS}, 1}(t^{-1})-\mc Z^+_{\text{HS}, 1}(t)+\sigma(t)
=(-1+6\,t^2+16\,t^3+\dots)+\sigma(t)$. The partition function here  starts from the contribution 
of the field strength $F_{\mu\nu}$  which  has dimension 2 and $\binom{4}{2}=6$ components and thus contributes  $6\,t^2$. The $-1$ term is spurious and is canceled by $\sigma_1 (t)=1$. 
This is, of course,   the $s=1$ case  of the general expression  for $\sigma_s (t)$  in footnote \ref{foot:sigma}.
Canceling spurious terms with a polynomial in $t+t^{-1}$ is in  general a convenient way to fix $\sigma(t)$.
A non-trivial example  is 4d  conformal graviton   where 
$\mc Z_2(t) = \mc Z^+_{\text{HS}, 2}(t^{-1})-\mc Z^+_{\text{HS}, 2}(t)+\sigma(t)
= (-{4}{t}^{-1} -7-4\,t+10\,t^2+40\,t^3+\dots) +\sigma(t)$. Here the first   contribution  to 
$\mc Z_2(t) $ should   be  from 
the Weyl tensor $C_{\mu\nu\rho\sigma}$ that has dimension 2 and 10 components. 
and thus  $\sigma_2(t) = 7+4\,(t+t^{-1})$  \cite{Beccaria:2014jxa}, 
  again in agreement with footnote \ref{foot:sigma}.
}
Thus, the expansion of  
$\log\text{I}_{\text{V}}$  should  contain the  term $-\sigma(1)\,\log\beta=-\log\beta$
(same as for  the 4d vector multiplet  in \rf{4.21}). 

The gauge field $B^-$  of the  tensor multiplet 
is dual to the  AdS$_7$ field with the  conformal representation 
content \cite{Beccaria:2015uta}
\be
\frac{1}{2}\Big[(4; 1,1,0)-(5; 1,0,0)+(6; 0,0,0)\Big],
\ee
where the prefactor $\frac{1}{2}$ takes into account the anti-selfduality constraint, cf. Table 1 of
\cite{Beccaria:2015uta}. Repeating the  steps leading to (\ref{six-2.1}), here we  find
\begin{align}
\mc Z_{B^-}(t) &= \mc Z^+_\text{HS}(t^{-1})-\mc Z^+_\text{HS}(t)+\sigma(t)
=\frac{1-6t+15t^2-15t^4+6t^5-t^6}{2\,(1-t)^6}+\sigma(t) \notag \\
&= (\frac{1}{2}+10\,t^3+45\,t^4+\dots)+\sigma(t).
\end{align}
Again, the first $t$-dependent term $10\,t^3$ is associated with the lowest dimension field which 
in this case is  the strength $H^-_{\mu\nu\rho}$ of $B^-$  (which   has  indeed   dimension 3
and $\frac{1}{2}\binom{6}{3}=10$ components). Then cancellation of the constant term 
 requires $\sigma(t)=-\frac{1}{2}$ and thus 
the expansion of $\log\text{I}_{\text{T}}$  should  contain 
the  term  $-\sigma(1)\,\log\beta=+\frac{1}{2}\log\beta$,  in agreement with  \rf{f8}.


\bibliography{BT-Biblio}
\bibliographystyle{JHEP}

\end{document}